\documentclass[traditabstract]{aa}  
\usepackage{natbib}
\usepackage{graphicx}
\usepackage{txfonts}
\usepackage{lscape}
\begin{document}

   \title{[O III] line properties in two samples of \\
radio-emitting narrow-line Seyfert 1 galaxies}


   \author{M. Berton
	\inst{1}\thanks{marco.berton.1@studenti.unipd.it}
	\and
	  L. Foschini\inst{2}
 	  \and
	S. Ciroi\inst{1}\and\\
	 V. Cracco\inst{1} 
	\and G. La Mura\inst{1}
	\and F. Di Mille\inst{3}
	\and P. Rafanelli\inst{1}	
          }

   \institute{$^{1}$ Dipartimento di Fisica e Astronomia "G. Galilei", Universit\`a di Padova, Vicolo dell'Osservatorio 3, 35122, Padova, Italy;\\
 $^{2}$ INAF - Osservatorio Astronomico di Brera, via E. Bianchi 46, 23807 Merate (LC), Italy;\\
	$^{3}$ Las Campanas Observatory, Carnegie Institution of Washington, Colina El Pino Casilla 601, La Serena, Chile.\\
             }

\authorrunning{Berton, M. et al.}
\titlerunning{[O III] lines in radio-emitting NLS1s}

\abstract{The [O III] $\lambda\lambda$ 4959,5007 lines are a useful proxy to test the kinematic of the narrow-line region (NLR) in active galactic nuclei (AGN). In AGN, and particularly in narrow-line Seyfert 1 galaxies (NLS1s) these lines often show few peculiar features, such as blue wings, often interpreted as outflowing component, and a shift $-$ typically toward lower wavelengths $-$ of the whole spectroscopic feature in some exceptional sources, the so-called blue outliers, which are often associated to strong winds. We investigated the incidence of these peculiarities in two samples of radio-emitting NLS1s, one radio-loud and one radio-quiet. We also studied a few correlations between the observational properties of the [O III] lines and those of the AGN. Our aim was to understand the difference between radio-quiet and radio-loud NLS1s, which may in turn provide useful information on the jet formation mechanism. We find that the NLR gas is much more perturbed in radio-loud than in radio-quiet NLS1s. In particular the NLR dynamics in $\gamma$-ray emitting NLS1s appears to be highly disturbed, and this might be a consequence of interaction with the relativistic jet. The less frequently perturbed NLR in radio-quiet NLS1s suggests instead that these sources likely do not harbor a fully developed relativistic jet. Nonetheless blue-outliers in radio-quiet NLS1s are observed, and we interpret them as a product of strong winds.}

\maketitle 
\newcommand{\prima}{$\lambda$4959}
\newcommand{\seconda}{$\lambda$5007}
\newcommand{\kms}{km s$^{-1}$}
\newcommand{\ergs}{erg s$^{-1}$}
\section{Introduction}
In the frame of the unified model \citep{Antonucci93}, narrow-line Seyfert 1 galaxies (NLS1s) are a class of active galactic nuclei (AGN) that exhibit somewhat unusual properties. Their main characteristics are the low full width at half maximum (FWHM) of the H$\beta$ line (by definition lower than 2000 km s$^{-1}$), and the low ratio [O III] $\lambda$5007\AA{}$/$H$\beta$ $<$ 3 \citep{Goodrich89,Osterbrock85}. Another important spectral feature is the Fe II multiplets, which are typical of type 1 AGN; this feature proves that we are directly observing the broad line region (BLR) despite the narrowness of the permitted lines. The low rotational velocity is commonly interpreted as the consequence of a relatively low mass central black hole ($10^{6-8}$ M$_\odot$) that is still growing fast. This has led to the conjecture that NLS1s might be young objects \citep{Grupe00, Mathur00} whose growth is driven by strong accretion onto the black hole, at rates that are closer to the Eddington limit than is the case for the other Seyfert classes \citep{Boroson92} and comparable to that of flat spectrum radio quasars (FSRQs; \citealp{Ghisellini10}). \par
Few of these sources (7\% according to \citealp{Komossa06}) are radio loud, meaning that the ratio between monochromatic fluxes (Jy) RL $= F_{5 \; GHz}/F_{B-band}$  is larger than 10, and they also exhibit some blazar-like properties, which are all signatures of a relativistic beamed jet \citep{Yuan08}. The presence of this jet was confirmed with the discovery by the \textit{Fermi Gamma-ray Space Telescope} of $\gamma$-ray emission from one of these radio-loud sources \citep{Abdo09a}, and was later followed by other detections \citep[see][]{Foschini15}. These $\gamma$-ray emitting, radio-loud NLS1s (RLNLS1s) have a flat radio spectrum and, in analogy with blazars, their misaligned counterparts should have a steep radio spectrum \citep{Urry95,Berton15a}. Steep-spectrum, radio-loud NLS1s do indeed exist, but their number is too low to represent the whole parent population of flat-spectrum radio-loud NLS1s \citep{Doi07,Doi11,Doi12,Gu10,Gliozzi10,Richards15}.\par
In order to solve this problem, \citet{Foschini11,Foschini12} suggested that the parent population might also include other kinds of sources beyond steep-spectrum radio-loud NLS1s. If the BLR has a flattened component, the Doppler effect can broaden the permitted lines when the inclination angle is high, making the source appear as a broad-line radio galaxy (BLRGs) or a narrow-line radio galaxy if the line of sight intercepts the molecular torus. \citet{Berton15a} investigated the problem, finding that the black hole mass distribution of BLRGs is partially overlapped with that of flat-spectrum NLS1s, suggesting a connection between these sources. Nevertheless there is another interesting possibility. If NLS1s are indeed young objects growing fast, the jet could have not developed radio lobes yet. In this case, it would be almost invisible for present days observatories, and the source would appear as a radio-quiet NLS1 (RQNLS1). In several works, jets were found in radio-quiet or mildly radio-loud NLS1s \citep[see][]{Doi15}; this could be a sign that even RQNLS1s might belong to the parent population and therefore appear, when observed pole-on, as flat-spectrum radio-loud NLS1s. \par
To investigate this problem, and in particular the relation between radio-quiet and radio-loud NLS1s, we focused on the properties of the [O III] $\lambda\lambda$ 4959,5007 lines. In AGN with high Eddington ratios, such as NLS1s, powerful outflows can be generated by the radiation pressure coming from the accretion disk \citep{Proga00}. The outflows have often been connected with the presence of an asymmetry in the [O III] lines \citep{Greene05}. These lines indeed show two distinct components. The first component is the line core, which typically has the same redshift as the whole galaxy. The second component usually has a higher FWHM than the first component and it is almost systematically blueshifted. This so-called blue wing (BW) has been directly associated with a gas outflow in the inner narrow line region (NLR). This is not the only peculiarity of the [O III] lines. In some cases, both of them show a blueshift with respect to their rest-frame wavelength. Those sources that exhibit this feature are known as blue outliers (BO). According to previous studies, they occur between $\sim$4\% and 16\% in NLS1s, depending on the definition \citep{Zamanov02,Komossa08}. \par
The generation mechanism of the [O III] shift is not well understood. A common hypothesis is that, as blue wings, the shift is induced by the strong winds generated by the high Eddington ratio. Nonetheless a different mechanism that can produce these blue outliers is a relativistic jet \citep{Tadhunter01,Komossa08,Nesvadba08}. Typically the NLR axis and the extended radio emission are aligned, a sign that a connection exists between these two features, and this is confirmed by the larger widths of narrow lines in AGN in which a nonthermal radio jet is harbored, possibly due to an acceleration of the gas by the relativistic plasma \citep{Pedlar85, Peterson}. The jet can indeed release part of its energy into thermal energy of the surrounding gas. The efficiency of this process is not well determined yet. Recent simulations \citep{Wagner11, Wagner12} showed that the efficiency is a function of the jet power, and that only powerful jets can affect the gas kinematics in the NLR, hence originating a blue outlier. \par 
The aim of this work is to investigate the incidence of BO and BW in two samples of radio-quiet and radio-loud NLS1s to determine whether a relativistic jet might be harbored by RQNLS1s. We investigate the [O III] line properties and the radio luminosity of the two samples, to understand if the mechanism powering the lines is the same. In Sect. 2 we describe the two samples of NLS1s, in Sect. 3 we describe our analysis procedure, in Sect. 4 we show our results, in Sect. 5 we discuss these results, and in Sect. 6 we briefly summarize our conclusions. Throughout this work, we adopt a standard $\Lambda$CDM cosmology, with a Hubble constant $H_0 = 70$ \kms\ Mpc$^{-1}$, and $\Omega_\Lambda = 0.73$ \citep{Komatsu11}. 
\begin{figure} 
\centering
\includegraphics[width=\hsize]{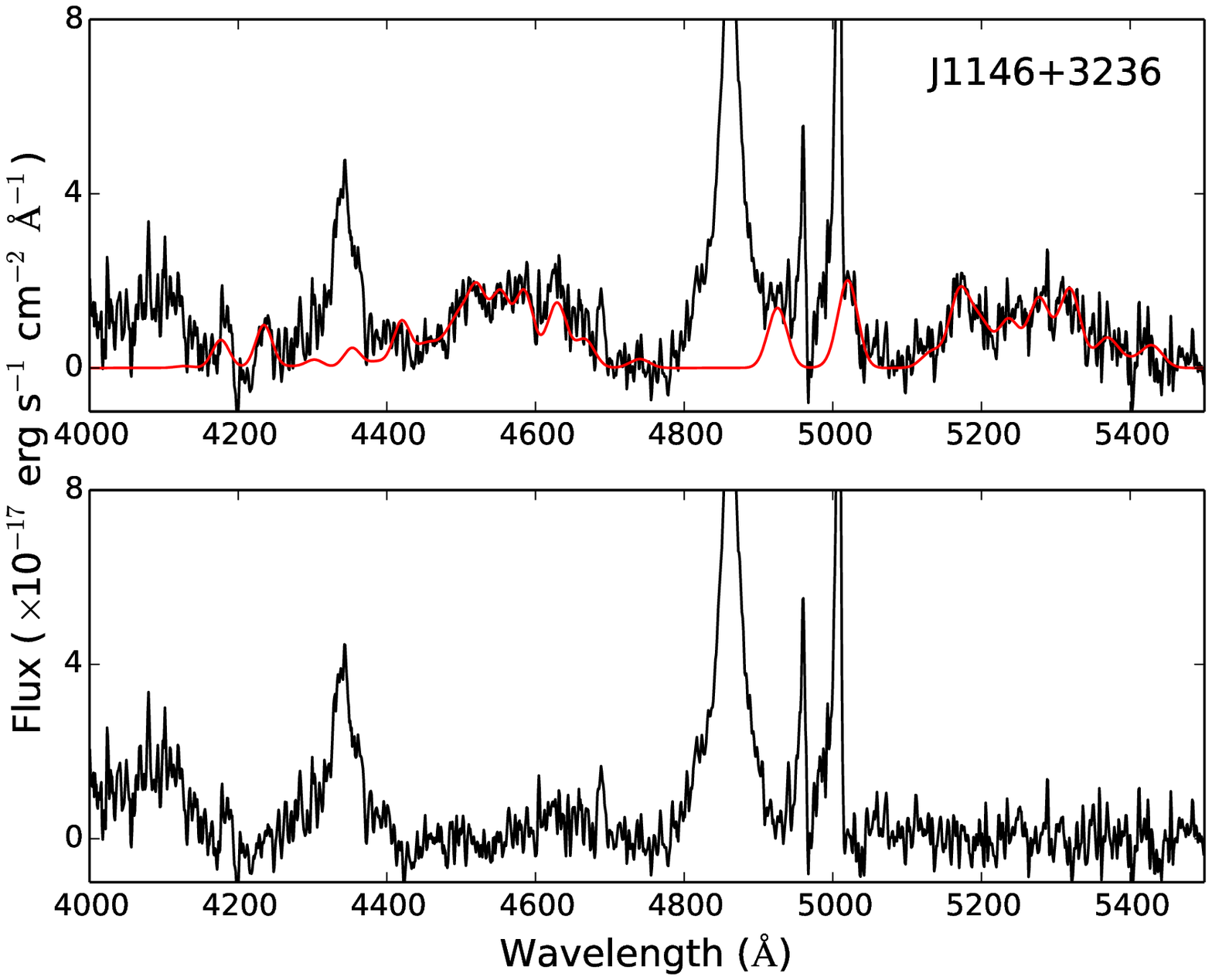}
\includegraphics[width=\hsize]{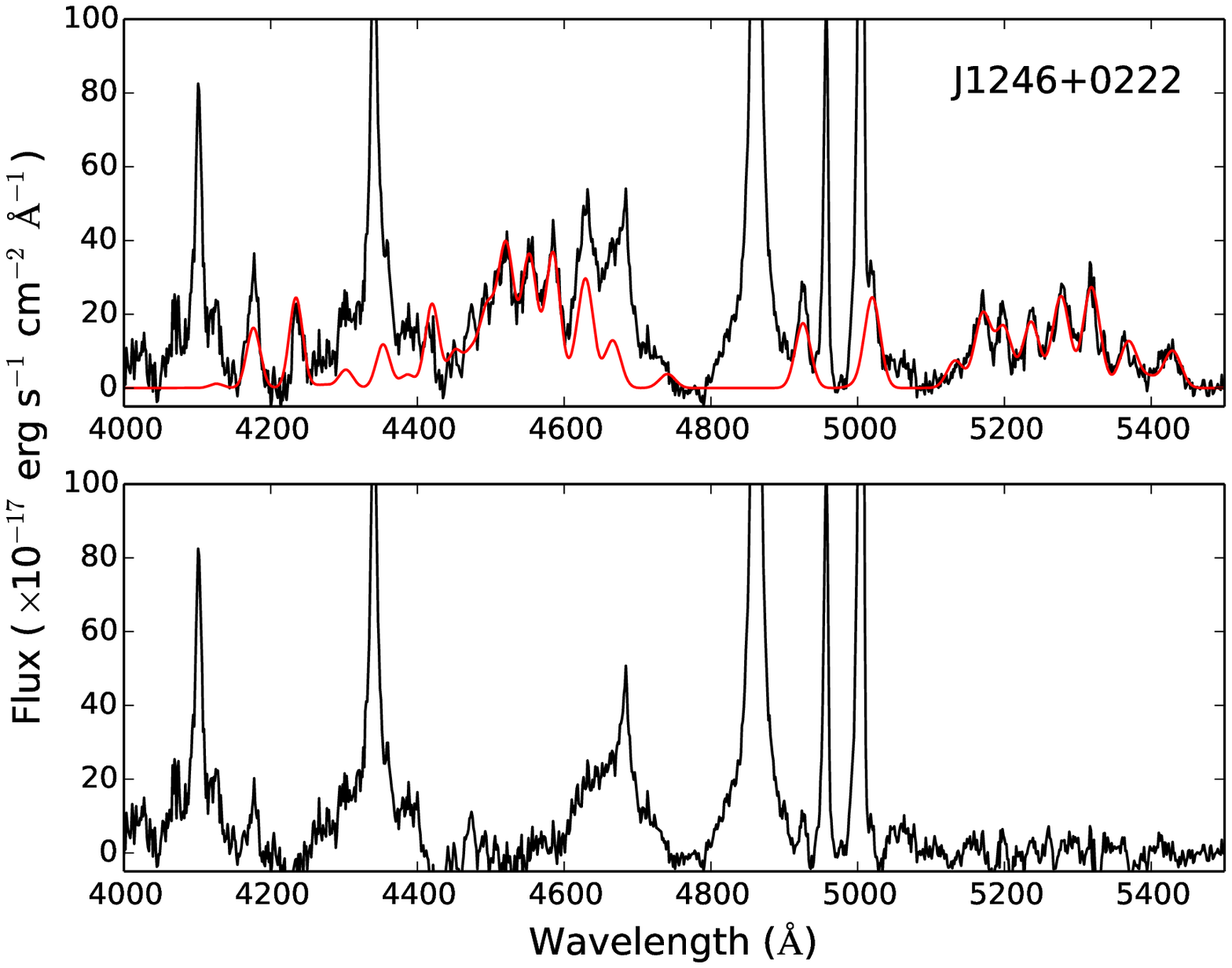}
\caption{H$\beta$ region of J1146+3236 (top 2 panels) and J1246+0222 (bottom 2 panels). The spectra have a S/N ratio of 10 and 40 in the 5100\AA{} continuum, respectively. In the first panel of each source the black solid line indicates the spectrum corrected for Galactic absorption, redshift and continuum subtracted; the red solid line indicates the Fe II template. In the second panel of each source the black solid line shows the spectrum with Fe II subtracted.}
\label{fit}
\end{figure}
\section{Samples selection}
\begin{figure}[t!]
\centering
\includegraphics[width=\hsize]{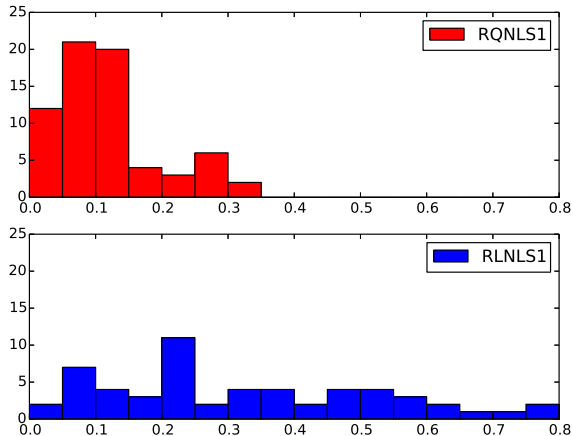} 
\caption{Redshift distribution of both samples. \textbf{Top panel:} radio-quiet NLS1s. \textbf{Bottom panel:} radio-loud NLS1s.}
\label{fig:redshift}
\end{figure}
\subsection{RQNLS1s}
We decided to use the sample created by \citet{Cracco16} to have a uniformly selected sample of RQNLS1s that is not contaminated by any spurious source. This sample was drawn from SDSS DR7 \citep{Abazajian09}, selecting only emission-line objects with a redshift between 0.02 and 0.35. These limits allowed us to stay within the spectral range both the [O II]$\lambda$3727 line and [S II]$\lambda\lambda$ 6717, 6731. They selected only sources with a FWHM of H$\beta$ between 800 and 3000 \kms. The lower limit is based on the measurements of type 2 and intermediate type AGN performed by \citet{Vaona12}. The upper limit is instead large enough to avoid loss of sources due to measurement errors. The final selection criterion was a signal to noise ratio (S/N) $>$ 3 in the [O I]$\lambda$ 6300 line. On the resulting sample these authors then applied the criteria of NLS1s, FWHM(H$\beta$) $<$ 2000 \kms\ and ratio [O III]/H$\beta <$ 3, obtaining 296 sources. The sample in this way is clean, containing only genuine NLS1s. They continued with a further selection by cross-matching the NLS1s sample with the FIRST survey \citep{Becker95}, and looking for radio sources within a radius of 5 arcsec from the SDSS coordinates. In this way they found 68 sources with an associated radio counterpart at 1.4 GHz. For each one they calculated the radio loudness. To obtain the B-band flux they convolved the spectra with a B-filter template, measuring the integrated-flux, while they assumed a spectral index of $\alpha_\nu = $ 0.5  to derive the 5 GHz flux (F$_\nu \propto \nu^{-\alpha_\nu}$, \citealp{Yuan08}). In this way they found 9 RLNLS1s, which we included in our second sample, and 59 RQNLS1s. To further increase the number of sources, we decided to add also the RQNLS1s from \citet{Berton15a}, which were not already included in their sample because of the selection criteria and had a suitable spectra to analyze. Our RQNLS1s sample in conclusion is made of 68 RQNLS1s. 
\subsubsection{RLNLS1s}
For the RLNLS1s sample, besides the nine sources we already found with the previous selection, we decided to use all the sources analyzed by \citet{Foschini15} and \citet{Berton15a} for which an optical spectrum was available in SDSS DR12, in the NED archive\footnote{http://ned.ipac.caltech.edu}, or observable with the Asiago 1.22m telescope (Sect.~\ref{sec:asiago}). All the objects were classified as NLS1s using the same spectral criteria specified before. Moreover, they all have a radio-loudness parameter RL $>$ 10. This sample includes both sources with a steep and flat radio spectrum, and also 26 without a measured spectral index. Steep-spectrum RLNLS1s are likely to be part of the parent population of flat-spectrum RLNLS1s \citep{Berton15a}, therefore they are the same kind of sources observed under a different angle. Our aim is to characterize RLNLS1s as a whole, so we decided to include all of them in our sample regardless of their spectral index. Our sample is comprised of 56 RLNLS1s. The redshift distributions of both samples is shown in Fig.~\ref{fig:redshift}. It is clear from the figure that the two distributions are very different. We accounted for this difference in the following.
\section{Data analysis}
\subsection{Data source}
\label{sec:asiago}
\begin{table}[t!]
\caption{Observational details for non-SDSS optical spectra. }
\label{tab:asiago}
\centering
\begin{tabular}{l c c c}
\hline\hline
Name & Exposure time (s) & R & Source \\
\hline
J0324+3410 & 3600 & 700 & A \\
J0632+6340 & 6100 & 2100 & T \\
J0706+3901 & 480 & 723 & N \\
J0713+3820 & 600 & 963 & N \\
J0806+7248 & 4800 & 700 & A \\
J0952+0136 & 4800 & 700 & A \\
J0925+5217 & 6000 & 1400 & A \\
J1203+4431 & 2400 & 1400 & A \\
J1218+2948 & 3600 & 1400 & A \\
J1337+2423 & 4800 & 700 & A \\
J1536+5433 & 2400 & 700 & A \\
J1555+1911 & 7200 & 1400 & A \\
\hline
\end{tabular}
\tablefoot{Columns: (1) short name; (2) exposure time in seconds; (3) spectral resolution R; (4) source of spectra: A for Asiago 1.22m telescope, T for Telescopio Nazionale Galileo, N for NED archive.}
\end{table}
We extracted 112 out of 124 optical spectra from the SDSS DR12. They have a resolution R $\sim$ 1700 and their wavelength calibration error is $\sim$2 \kms\ \citep{Abazajian09}. Five radio-quiet sources and one radio-loud were not included in SDSS, so we observed them with the Asiago 1.22m telescope. Moreover, three sources, J1218$+$2948, J1555$+$1911, and J1337$+$2423, are in the SDSS archive, but in the first two cases the optical spectra were not taken on nucleus so they show just the host galaxy contribution, while the last one was taken outside the galaxy, and its spectrum is just pure noise. For this reason, we decided to reobserve them with the Asiago telescope. These nine spectra were obtained between 2014 January and 2015 March, using a Boller \& Chivens spectrograph with a 300 mm$^{-1}$ grating. The spectral resolution is between $\sim$700 and $\sim$1400, depending on the seeing conditions. We divided our observations in frames of 1200 s each, to decrease the contamination from cosmic rays and light pollution. All of these spectra were reduced using the standard \texttt{IRAF v.2.14.1} tasks, using HeFeAr lamp for wavelength calibration and overscan in place of bias. The wavelength calibration error, evaluated on the HeFeAr lamp, is on average $\sim$20 \kms. Two more spectra were derived from the NED archive. Finally, one was obtained in October 2005 using the 3.58 m Telescopio Nazionale \textit{Galileo} (TNG) with the DOLORES camera (device optimized for the low resolution). We used the MR-B Grm2 grism and a slit of 1.1" with an He lamp for wavelength calibration. Sources for non-SDSS spectra, exposure times, and resolution are summarized in Tab.~\ref{tab:asiago}.\par
\subsection{Preliminary correction}
All the spectra were corrected for Galactic absorption, using the column density values reported in \citet{Kalberla05}. We then corrected for redshift. According to \citet{Komossa08}, the best method to perform this last procedure is to use stellar absorptions lines $-$ that are not visible in our NLS1s spectra $-$ as reference or, alternatively, to use low ionization lines as [S II]$\lambda\lambda$6716,6731 and [O II] $\lambda$3727. Nevertheless [S II] is not visible in the spectra of RLNLS1s when $z \gtrsim 0.35$. The [O II] is instead present in 113 spectra, while it is not visible in the remaining six radio-loud and five radio-quiet. We decided to use it as reference. This line is actually a doublet and the two lines are not resolved in any of our spectra. The ratio between the doublet lines is a function of gas density. The position of the line is then a function of both the systemic velocity and density. Nevertheless, the density cannot be estimated in all of our spectra. Therefore it is not possible to quantify this error, which induces a small scatter in our measurements. This problem was already pointed out by \citet{Boroson05}, who concluded that this doublet can be used as reference for the systemic velocity of the galaxy, although in few cases it might introduce some bias. \par
The line was fitted with a single Gaussian using an automatic procedure developed in \texttt{Python}. To compute the error on its position we used a Monte Carlo method. It consists in repeating the measurement 100 times at the same time varying the flux with a random Gaussian noise proportional to the rms measured in the 5100 \AA{} continuum. This procedure was used in the following to evaluate all the errors on the lines. Since the fit was performed with only one Gaussian, the typical error associated with our measurement is low, and on the same order of the wavelength calibration of SDSS spectra, $\sim$ 2 \kms. A larger systematic error is instead induced by the doublet nature of the [O II] line we already mentioned. \citet{Boroson05} calculated that the standard deviation of its redshift distribution calculated with respect to other high ionization lines is 39 \kms, even if this value is likely an upper limit due to some other systematic errors, and its actual value is likely lower. Despite this relatively large systematic uncertainty, however, in our case the [O II] remains the most reliable option. \par
In those cases where the [O II] line was not present, we used the narrow component of H$\beta$ line (H$\beta_n$) to determine the redshift, as in \citet{Zamanov02} and \citet{Marziani03}. We reproduced the line profile using an automatic procedure which performs the fit using alternatively two or three Gaussians, one for the narrow and one or two for the broad component. The number of Gaussians was decided by the software according to the reduced chi-squared, $\chi^2_\nu$, of the fit. We did not use any constrain on position, intensity or width of the components. The typical error in the H$\beta_n$ position, calculated again with the previously described Monte Carlo method, is $\sim$0.1 \AA{} for two Gaussians, and $\sim$0.5 \AA{} for three Gaussians, corresponding to $\sim$ 10 and $\sim$30 \kms, respectively. Because of these relatively large uncertainties, we decided not to use the H$\beta$ line as reference in all of our sources, but only when the [O II] was not present.\par

\subsection{Fe II subtraction}
After these preliminary corrections, we focused on the [O III] region fitting, where we removed the power-law continuum of the AGN. We neglected the host galaxy component, since all the spectra are clearly dominated by the AGN, showing no signs of stellar contribution. We then removed the Fe II multiplets using the online software\footnote{http://servo.aob.rs/FeII\_AGN/} developed by \citet{Kovacevic10} and \citet{Shapovalova12}. This template reproduces 65 FeII emission lines between 4000 and 5500\AA{}, divided into five line groups, fitting each line with a single Gaussian. The required input parameters are Doppler width, shift in velocity of the Gaussians, intensity of each group of multiplets, and excitation temperature. We chose the input parameters after a preliminary measure on the spectra. For the Doppler width, in particular, we used as a first approximation the FWHM of H$\beta$, since both Fe II and H$\beta$ are emitted mostly within the BLR. The online software creates a model that can be directly subtracted to the spectrum. The quality of the result was estimated by checking whether the residuals were comparable to the noise of the spectrum. The flux error, estimated by measuring the same model with different values of noise, is about $\sim$10\% for a S/N = 20 in the 5100\AA{} continuum, and $\sim$20\% for a S/N = 10. Typical values of s/N are $\sim$15 for RLNLS1s and above $\sim$20 for RQNLS1s. We can then assume an average error of $\sim$15\% for RLNLS1s and of less than 10\% for RQNLS1s. An example of Fe II subtraction is shown in Fig.~\ref{fit}. \par

\subsection{H$\beta$ line}
After the Fe II subtraction we continued with the fitting of H$\beta$. This line is crucial for evaluate the properties of the central engine in each NLS1s. As mentioned before, we decomposed it using alternatively two or three Gaussians as previously described. We then calculated the black hole mass, the bolometric luminosity and Eddington ratio for each source. To do this, we followed the steps described by \citet{Berton15a}. Our calculations are performed under the hypothesis of virialized system. As a proxy for the rotational velocity, we used the second-order moment of broad H$\beta$ instead that the FWHM because it is believed to be less affected by inclination and BLR geometry \citep{Peterson04, Peterson11}. To estimate the BLR radius, we exploited its relation with H$\beta$ luminosity obtained by \citet{Greene10}. Further details can be found in Sect.~4 of \citet{Berton15a}. Our results, shown in Tables~\ref{tab:summary1} and \ref{tab:summary2}, are in very good agreement with those found in \citet{Foschini15} and \citet{Berton15a}.\par
\begin{figure}[t!]
\centering
\includegraphics[width=\hsize]{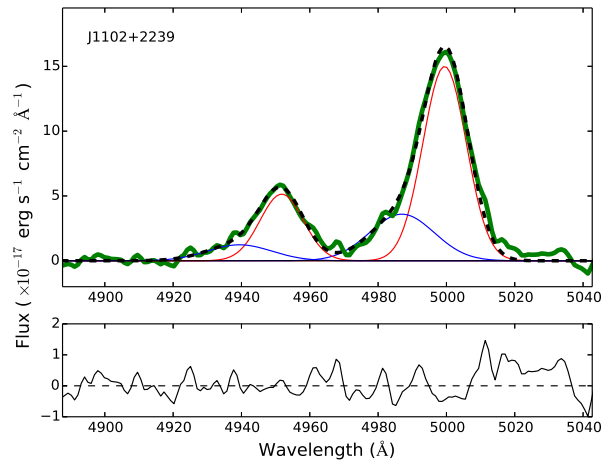} 
\caption{[O III] lines fitting with the automatic procedure in J1102+2239. \textbf{Top panel:} the thick green line is the original spectrum, continuum and Fe II subtracted. The blue and red solid lines represent the blue wing and core component, respectively. The black dashed line is the sum of the resulting fit. \textbf{Bottom panel:} residuals of the fitting procedure.}
\label{J1102}
\end{figure}
\subsection{[O III] lines}
The final step is the [O III] lines fitting. To reproduce their profile we used four Gaussians, that is, two for each line. The first Gaussian is the blue wing, while the second is the core component. In all cases we fixed the flux ratio between each component of the \prima\ and \seconda\ lines using its theoretical value of 1/3. We used the \prima\ line to verify the validity of the fit in the \seconda\ line. When \prima\ had an amplitude lower than three times the RMS of the continuum at 5100 \AA, we fitted just the \seconda\ line. The fitting procedure was performed using an automatic procedure that also allows us to estimate the errors on each parameter using the Monte Carlo method again. An example of [O III] lines automatic fitting is shown in Fig.~\ref{J1102}. \par
In 14 sources, 11 radio-loud and three radio-quiet, we were not able to fit the [O III] lines with both the core and the wing. This occurs when the line has a S/N (four cases, all radio-loud) that is too low, or when it was already well reproduced by a single Gaussian (10 cases, including seven radio-loud and three radio-quiet). In those cases we only measured the peak position of the line, its core width, and its flux. We measured the peak wavelength of each Gaussian component, its FWHM, and the total flux of the \seconda\ line. An example of fit is shown in Fig.\ref{fit}. Finally all the FWHM values were corrected for instrumental resolution, which was $\sim$167 \kms\ for SDSS spectra, and that is specified in Tab.~\ref{tab:asiago} for the other spectra. All of our results and their errors are shown in Tables~\ref{tab:summary3} and \ref{tab:summary4} for radio-quiet and radio-loud sources, respectively.\par
\begin{figure}[t!]
\centering
\includegraphics[width=\hsize]{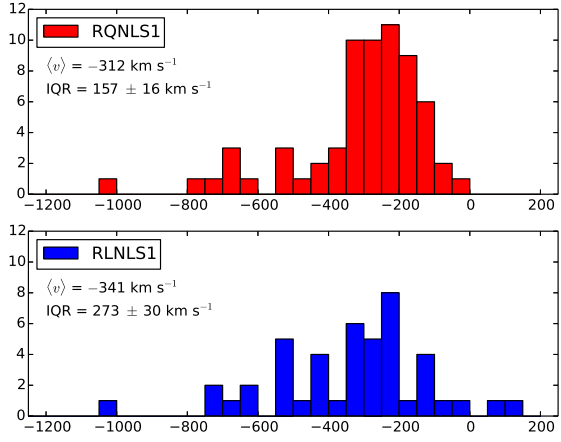} 
\caption{Histogram showing the velocity shift of the blue wings with respect to the core component. The negative velocity is due to the approaching gas. In top panel the RQNLS1s sample is shown; in bottom panel the RLNLS1s sample is shown. In each panel, the average velocity, IQR, and its standard deviation are shown.}
\label{bluewings}
\end{figure}
\section{Results}
\begin{table}[t!]
\caption{Correlation coefficients between the wing velocity and other quantities.}
\label{correlazioni}
\centering
\begin{tabular}{l c c c }
\hline\hline
Sample & Q1 & $r$ & p-value \\
\hline
RQ non-outliers & FWHM$_c$ & -0.5 & 8$\times10^{-6}$ \\
RL non-outliers & FWHM$_c$ & -0.5 & 1$\times10^{-3}$ \\
RQ+RL non-outliers & FWHM$_c$ & -0.5 & 2$\times10^{-8}$ \\
RQ outliers & FWHM$_c$ & 0.4 & 6$\times10^{-1}$ \\ 
RL outliers & FWHM$_c$ & -0.2 & 6$\times10^{-1}$ \\ 
RQ+RL outliers & FWHM$_c$ & 0.1 & 8$\times10^{-1}$ \\
RQ non-outliers & FWHM$_w$ & -0.2 & 8$\times10^{-2}$ \\
RL non-outliers & FWHM$_w$ & -0.2 & 2$\times10^{-1}$ \\
RQ+RL non-outliers & FWHM$_w$ & -0.2 & 2$\times10^{-2}$ \\
RQ outliers & FWHM$_w$ & 0.5 & 5$\times10^{-1}$ \\ 
RL outliers & FWHM$_w$ & 0.1 & 9$\times10^{-1}$ \\ 
RQ+RL outliers & FWHM$_w$ & 0.2 & 4$\times10^{-1}$ \\
RQ & Eddington & -0.1 & 7$\times10^{-1}$ \\
RL & Eddington & -0.2 & 3$\times10^{-1}$ \\
RQ & L$_{bol}$ & -0.3 & 2$\times10^{-2}$ \\
RL & L$_{bol}$ & -0.3 & 3$\times10^{-2}$ \\
RQ & L$_{rad}$ & -0.3 & 2$\times10^{-2}$ \\
RL & L$_{rad}$ & -0.1 & 7$\times10^{-1}$ \\
RQ & M$_{BH}$ & -0.3 & 7$\times10^{-3}$ \\
RL & M$_{BH}$ & -0.2 & 1$\times10^{-1}$ \\
\hline
\end{tabular}
\tablefoot{Columns: (1) tested sample; (2) tested quantity ; (3) Pearson $r$ coefficient; (4) Pearson p-value.}
\end{table}

\begin{figure}
\centering
\includegraphics[width=\hsize]{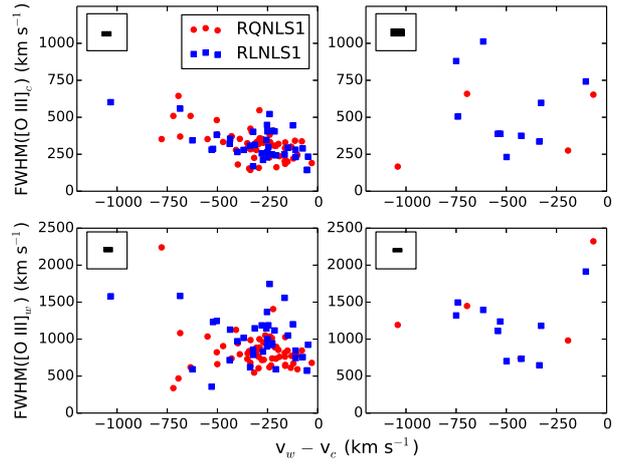} 
\caption{In top panels the velocity of the wing against FWHM of the core component for both samples, regular sources in the left and outliers in the right. In bottom panels the velocity of the wing against FWHM of the wing for both samples, as before. RQNLS1s are indicated by red circles, RLNLS1s are indicated by blue squares. The black filled rectangles represented in the boxes at the upper left corner of each diagram illustrate the size of the average uncertainty range of our measurements.}
\label{core-wing}
\end{figure}
\subsection{Blue wings}
The blue wing is represented by the lower wavelength peaked Gaussian in both the [O III] lines. In four radio-loud sources neither the \prima\ nor the \seconda\ lines could be modeled with two Gaussians because the lines were too weak, while in seven radio-loud and three radio-quiet sources the [O III] appears to have only a core component. In all the other cases we decomposed one or both the lines as previously explained, finding a blue wing in 108/110 cases, and a red wing only in two sources that were both radio-loud. \par
Fig.~\ref{bluewings} shows that the two distributions of the blue wing velocity are pretty similar. To compare these distributions we performed an Anderson-Darling (A-D) test \citep{Hou09}. The null hypothesis is that the two distributions are drawn by the same population. The rejection of the null hypothesis throughout this work is fixed to a p-value below 0.05. The resulting p-value of the test is 0.33, therefore the null hypothesis cannot be rejected. Their average velocities, also indicated in Fig.~\ref{bluewings}, are quite close, with a slightly higher value among RLNLS1s. The interquartile range (IQR) of the radio-loud sample, however, is almost double that of RQNLS1s. We evaluated the uncertainty on this quantity using another Monte Carlo simulation. We varied each blue wing velocity using a Gaussian noise, whose width was equal to the error on each measurement. Then we calculated the IQR of the new distribution and we repeated this procedure 1000 times to compute the standard deviation of the IQR. The results are shown in Fig.~\ref{bluewings} and they indicate that the velocity distribution in the radio-loud sample has a larger intrinsic scatter. \par
We tested the correlation between the velocity of the wing and the FWHM of the core and wing component, respectively. The results are summarized in Tab.~\ref{correlazioni}. Those sources where it was not possible to separate core and wing are not considered. At the beginning we did not find any correlation between these quantities. We then decided to test these correlations for regular sources and outliers separately; these results are shown in Fig.~\ref{core-wing}. As in \citet{Xiao11}, we found a moderate but significant anticorrelation (Pearson $r$ = -0.5, p-value = 4$\times$10$^{-8}$) between the core component of [O III] and the wing velocity in the nonoutliers sources, while we found no correlation ($r$ = 0.1, p-value = 8$\times$10$^{-1}$) among the outliers. Conversely we did not find a correlation between the wing FWHM and its velocity. These results, in particular the correlation among core FWHM and wing velocity, might be explained if a gas where a turbulent outflow is generated is turbulent itself. This gas would then show a high core FWHM due to this turbulence. \par
\begin{figure}[t!]
\centering
\includegraphics[width=\hsize]{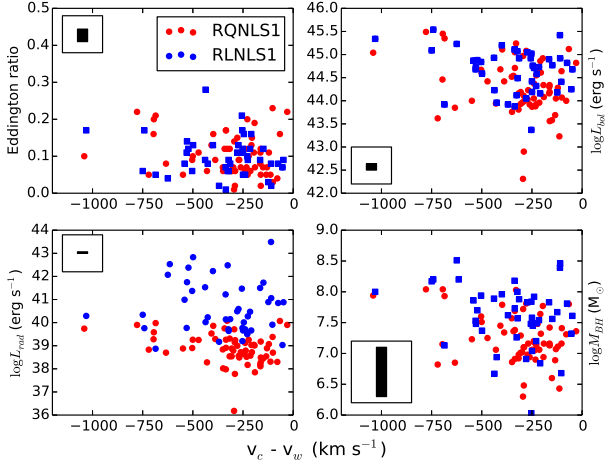} 
\caption{Correlations between the absolute value of the [O III] wing component velocity, in abscissa, and other quantities. \textbf{Top left:} Eddington ratio; \textbf{top right:} logarithm of the bolometric luminosity (\ergs); \textbf{bottom left:} logarithm of radio luminosity at 1.4 GHz (\ergs); \textbf{and bottom right:} logarithm of the black hole mass (M$_\odot$). RQNLS1s are indicated by red circles, RLNLS1s are indicated by blue squares. Black filled rectangles as in Fig.~\ref{core-wing}.}
\label{bw-edd}
\end{figure}

 
The blue wings are thought to originate in outflows induced by the high Eddington ratio \citep{Whittle85, Komossa08}. For this reason we looked for a correlation between the velocity of the wing and the Eddington ratio. We also tested the correlation with the black hole mass and the bolometric luminosity, since these quantities are directly related to the Eddington ratio and the radio luminosity which, if a jet is present, may have some effect on the gas velocity. The results are shown in Fig.~\ref{bw-edd} and it is already evident even by visual inspection that no correlation is present among these quantities. The only significant, but very weak, trend is only between the wing velocity and the black hole mass in RQNLS1s ($r$ = 0.3, p-value = 7$\times10^{-3}$). In this case blue outliers and regular sources seem to behave in the same way, since the exclusion of outliers does not change our results significantly.
\subsection{Blue outliers}
\begin{figure}[t!]
\centering
\includegraphics[width=\hsize]{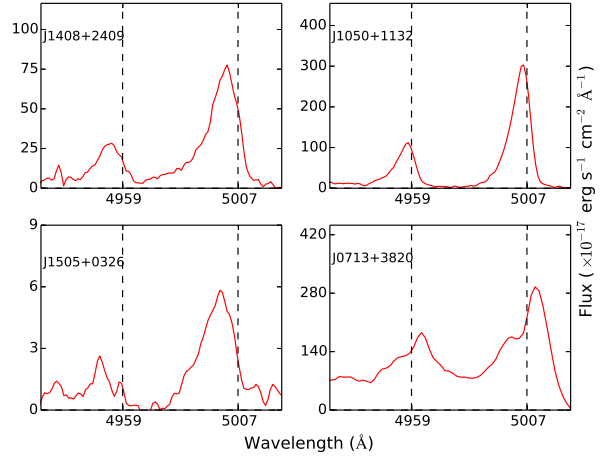} 
\caption{Four examples of blue and red outliers in our samples, continuum and Fe II subtracted. The vertical dashed lines are the restframe position of [O III] \prima\ and \seconda. Those in the two upper panels are RQNLS1s, while those in the bottom panels are RLNLS1s.}
\label{multiplot}
\end{figure}
\begin{figure}[t!]
\centering
\includegraphics[width=\hsize]{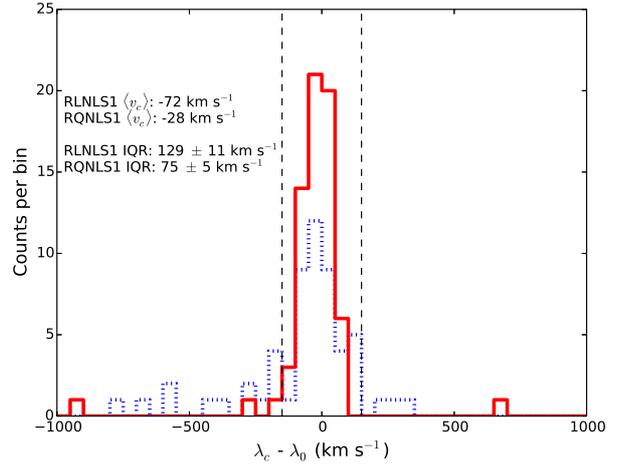} 
\caption{Histogram of the peak position of the \seconda\ line in the two samples with respect to [O II]. Binning of 50 \kms. The dashed vertical lines are the limits of 150 \kms\ for blue and red outliers. The red solid line indicates the RQNLS1s sample; the blue dotted line indicates the RLNLS1s sample. The values are the IQR, average shift, and standard deviation, all in \kms.}
\label{blueoutliers}
\end{figure}

As mentioned in the introduction, few NLS1s show a blueshift of the [O III] lines. These kinds of sources are called blue outliers and a few examples are shown in Fig.~\ref{multiplot}. To look for them we calculated the distance between the measured peak of the \seconda\ core component and its rest-frame wavelength (5006.843 \AA{}), converted into velocity. As in \citet{Komossa08}, we defined an object as a blue outlier when this velocity is $v_{[O III]} \leq -150$ \kms. We also defined an object as red outlier when the [O III] line is shifted toward higher wavelengths of the same quantity. Fig.~\ref{blueoutliers} shows the distributions of the \seconda\ line velocity shift, the IQR and the average shift values for each sample. The larger IQR in the radio-loud velocity distribution is strictly connected with the number of outliers. In the radio-quiet sample we found only one red and three blue outliers ($\sim$6\%), while in the radio-loud there are 13 blue outliers and three red outliers ($\sim$29\%). \par 
\begin{figure}[t!]
\centering
\includegraphics[width=\hsize]{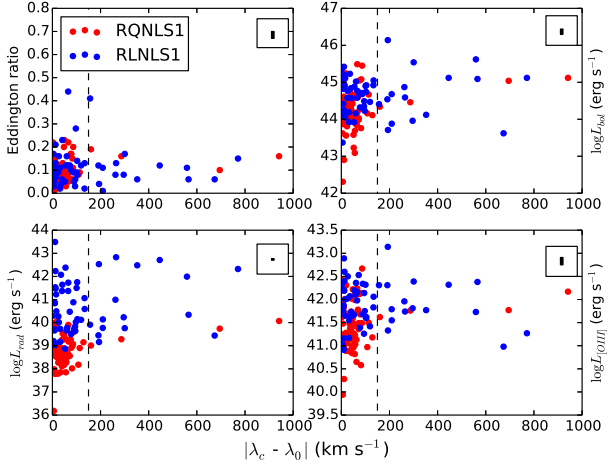} 
\caption{Correlations between the absolute value of [O III] velocity shift, in abscissa, and other quantities. \textbf{Top left:} Eddington ratio; \textbf{top right:} logarithm of the bolometric luminosity (\ergs); \textbf{bottom left:} logarithm of radio luminosity at 1.4 GHz (\ergs); \textbf{and bottom right:} logarithm of the [O III] luminosity. The dashed vertical line indicates the limit for outliers. RQNLS1s are shown with red circles, RLNLS1s are shown with blue squares. Black filled rectangles as in Fig.~\ref{core-wing}.}
\label{bo-edd}
\end{figure}
In both the radio-quiet and radio-loud samples the majority of sources have a typical velocity between -50 and 50 \kms (41/68 in RQNLS1s, 21/56 in RLNLS1). The RLNLS1s show a few more redshifted sources, and in general they appear to be distributed over a larger interval of velocities. This is shown by the IQR of the distributions, which is 129 $\pm$ 11 \kms\ for RLNLS1s and 75 $\pm$ 5 \kms\ in RQNLS1s. Hence, RQNLS1s have a narrower distribution, and this suggests that the gas is more perturbed in RLNLS1s. We performed an A-D test, finding a p-value of 0.04 which allows us to reject the null hypothesis. This result can be interpreted as a sign that the two distributions are originated via different mechanisms.\par
In our study of outliers the measurement errors must be considered. As mentioned before, we calculated the error on both the [O III] line core position and the [O II] line via Monte Carlo method. By considering both of these errors, all four sources in the RQNLS1s sample are still outliers. In RLNLS1s instead the numbers vary between 13 (11 blue, 2 red) and 17 (13-4). The number of outliers is then systematically larger in the radio-loud sample. It is worth noting that the null hypothesis of the A-D test cannot be rejected when only 13 outliers are present in the radio-loud sample (p-value 0.15). Conversely when 17 outliers are considered in the radio-loud sample, the null hypothesis is rejected with a higher confidence level (p-value 4$\times$10$^{-3}$).\par
To obtain further confirmation of our results, we measured the incidence of blue outliers in all the sources not detected in the FIRST survey from the \citet{Cracco16} sample we already described. This sample includes 227 sources, and we found that the number of blue outliers is between three and five, depending on the errors. The A-D test, performed between the radio-loud sample and the radio-quiet+radio-silent sample, allows us to reject the null hypothesis even in the worst case, with a p-value of 6$\times10^{-3}$. This result is particularly significant because it reveals a strong correlation between the strength of the radio emission and number of outliers. A strong turbulence in the NLR then is often, even if not always, associated with a high radio luminosity. \par
We finally investigated the presence of a correlation between the blue outliers and the Eddington ratio, to understand whether the shift of [O III] is also connected with the high accretion rate of NLS1s. This correlation was first found by \citet{Marziani03} and later confirmed by \citet{Bian05}. Nevertheless, in agreement with \citet{Aoki05}, we did not find any correlation between the Eddington ratio and the blue outliers (see Fig.~\ref{bo-edd}). We also tested the correlation between the blue outliers and other significant quantities that might in some way affect the gas kinematics, such as the bolometric luminosity, radio luminosity, and [O III] luminosity. In particular, a high bolometric luminosity might affect the gas by means of the radiation pressure. In a similar way, the radio luminosity is linked with the relativistic jets properties. If the jet is present they might be connected in some way with the outliers velocity. Finally, a high [O III] luminosity can be connected to gas dynamics. Again, however, we did not find any significant result, where the lowest p-value is 0.01 in RQNLS1s between the core shift and radio luminosity ($r$ = 0.3). \par
We investigated the same correlations considering only the outliers, and we found that, while the radio-loud outliers have the same behavior as the whole sample, radio-quiet outliers seem to follow a different trend, although the statistic is very sparse. In particular, we found a very strong and significant correlation between the radio luminosity and the [O III] core shift ($r$ = 0.99, p-value = 4$\times$10$^{-3}$). An equally strong correlation, although less significant, can be found with the bolometric luminosity ($r$ = 0.98, p-value = 0.02). This correlation is not present in RLNLS1s ($r$ = 0.3, p-value = 0.32). It is worth noting that in both samples no outliers can be found below a radio luminosity of 10$^{39}$ \ergs, even if 56\% of radio-quiet sources lies below that threshold. This result seems to hold even for two out of three radio-silent sources, which indeed have an upper limit of radio luminosity above this threshold. A high radio luminosity might then be very important in the production of outliers.
\subsection{Radio versus [O III]}
\begin{figure}[t!]
\centering
\includegraphics[width=\hsize]{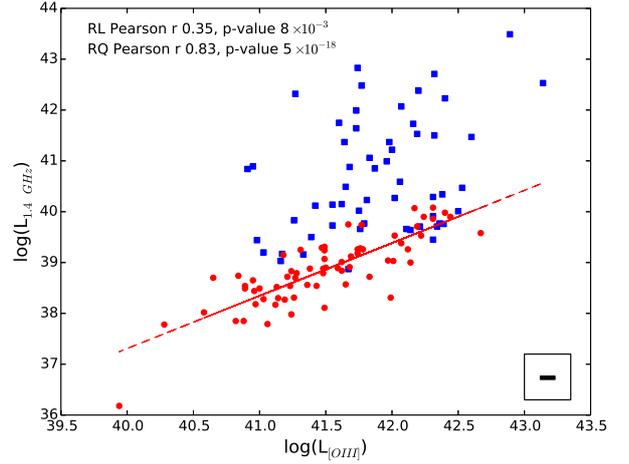} 
\caption{[O III] luminosity vs. radio luminosity at 1.4 GHz. The solid red line indicates the best fit for RQNLS1s. Correlation coefficients for both samples are also shown. Black filled rectangles as in Fig.~\ref{core-wing}. }
\label{radioo3}
\end{figure}
As done for the first time by \citet{deBruyn78}, we searched for a correlation between the luminosity of [O III] and the radio luminosity at 1.4 GHz extracted from FIRST survey. The result is shown in Fig.\ref{radioo3}. It is evident that RQ and RL sources form two distinct populations in the plot. RQNLS1s show a strong correlation between these two quantities ($r$ = 0.8, p-value = 5$\times$10$^{-18}$), while in RLNLS1s there is only a trend ($r$ = 0.4, p-value = 8$\times10^{-3}$). The lack of a strong correlation among radio-loud sources is due to very large scatter in radio luminosity. We cannot rule out that, once considered the sources with a radio emission too faint to be detected, a similar scatter would be present among radio-quiet sources too. The detection limit, because of the redshift, can induce a spurious correlation between the radio and [O III] luminosity, since sources with a high [O III] luminosity but a weak radio emission cannot be detected. To test this possibility, we used the software developed by \citet{Akritas96}, which allows us to test the existence of a correlation in presence of censored data. In this way we found that the null hypothesis of no correlation can be rejected at a confidence level $>$ 5\%. Therefore the correlation among these luminosities is present even when the redshift effects are considered. \par
The linear best-fit relation for the radio-quiet sample can be expressed as 
\begin{equation} 
\log{L_{1.4 \, GHz}} = (-4.18\pm0.47) + (1.04 \pm 0.01)\log{L_{[O III]}} \; ,
\label{eqo3}
\end{equation}
with a scatter of 0.2 \textit{dex}. It is worth noting that some radio-loud sources seem to lie close to this relation and are completely overlapped with the radio-quiet sample. \par
The [O III] luminosity interval in which our samples are located are different, as confirmed by a Kolmogorov-Smirnov test (K-S, p-value = 2$\times$10$^{-4}$). This difference is likely to be a selection effect owing to the different redshift distributions of our samples. Fig.~\ref{radioo3} shows how those sources with high radio luminosity, hence visible at high $z$, also have a high [O III] luminosity. Such a line then is particularly bright in radio-loud sources. In NLS1s the [O III] line flux must be, by definition, on the same order of magnitude of H$\beta$, so this line must be equally bright. These strong optical lines allow an easier classification of NLS1s. Radio-loud NLS1s are more likely to be identified as such in a large survey. This likely induces a selection effect. We indeed expect that at high redshift the relative number of detected RLNLS1s with respect to RQNLS1s increases for this very reason. \par 
It must be highlighted that the number of blue and red outliers might be larger in RLNLS1s because they are typically at larger redshift. The difference between the samples would therefore be due to evolution, and not to the relativistic jet. Nonetheless, if this was true, among RLNLS1s we should observe outliers only at high redshift, but exactly half of the outliers are located below z=0.35, which is the upper limit for our radio-quiet sample. Hence the redshift seems not to have a significant incidence on the outliers presence. The same is true for bolometric luminosity and [O III] luminosity, neither of which, as shown before, are correlated with the shift of [O III], therefore our results might be interpreted as a true physical difference in the NLR of our two samples.\par
\begin{figure}[t!]
\centering
\includegraphics[width=\hsize]{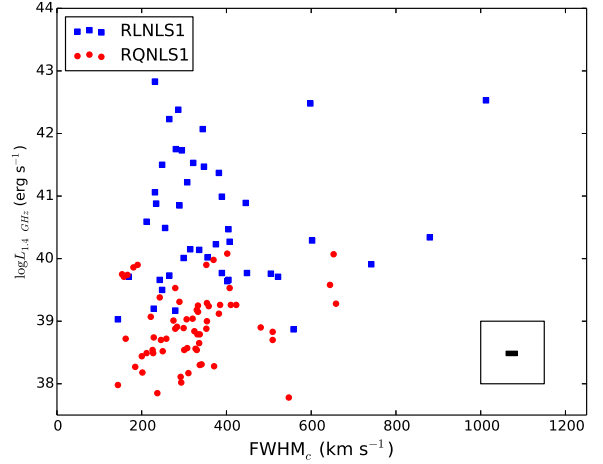} 
\caption{FWHM of [O III]$_c$ (in \kms) against radio luminosity at 1.4 GHz (in erg s$^{-1}$). RQNLS1s are indicated by red circles, RLNLS1s are indicated by blue squares. Black filled rectangles as in Fig.~\ref{core-wing}.}
\label{core-radio}
\end{figure}

As found for the first time by \citet{Pedlar85}, and recently confirmed by \citet{Mullaney13}, the radio-luminosity has an effect on the [O III] line profile. Therefore we searched for a correlation between the FWHM of the core component of [O III] and the radio luminosity at 1.4 GHz, neglecting those sources in which we could not separate the core and wing components. We did not find such correlation, as shown in Fig.~\ref{core-radio}, with $r$ = 0.0, p-value = 0.8 in RLNLS1s, and $r$ = 0.2 and p-value = 0.2 among RQNLS1s. 
 \par

\subsection{Ratio [O III]/H$\beta$}
\begin{figure}[t!]
\centering
\includegraphics[width=\hsize]{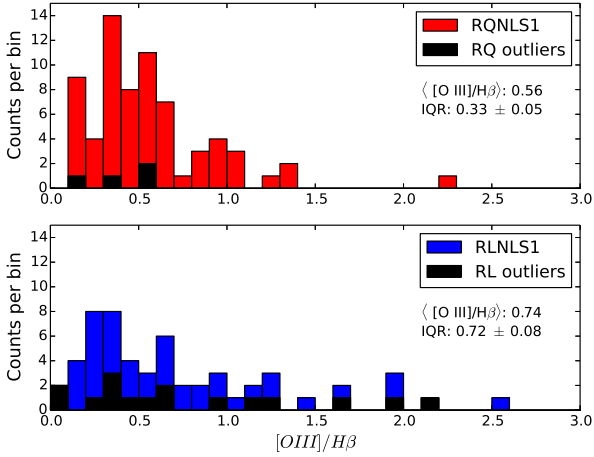} 
\caption{Histogram of the distribution of ratio R5007 between [O III] and H$\beta$. The ratio for blue outliers is indicated by the black histogram. In the top panel, the RQNLS1s sample is shown; in the bottom panel, the RLNLS1s sample is shown. Binning of 0.1.}
\label{o3hb}
\end{figure}

We finally measured the ratio between the [O III] \seconda\ line and the whole H$\beta$ flux, also known as the R5007 parameter. Since the broad H$\beta$ line is formed in the inner part of the BLR \citep{Greene05b}, this ratio is a useful tool to evaluate whether the jet/gas interaction is different in BLR and NLR. The histogram with the results is shown in Fig.~\ref{o3hb}. The average of this ratio for radio-quiet sources is 0.56 with an IQR of 0.33$\pm$0.05, while for radio-loud sources the average is 0.74 with IQR 0.72$\pm$0.08. The difference is not significant, in fact both the K-S and the A-D tests do not reject the null hypothesis (p-values 0.23 and 0.09, respectively). We also investigated the ratio in the blue outliers, also shown in Fig.~\ref{o3hb}. In the radio-quiet sample, all the sources have a R5007 below 0.6, that is systematically located in the low ratio region of the histogram, while among radio-loud sources the results are distributed over a larger interval. The mean ratio for radio-quiet blue outliers is 0.42 with IQR 0.24$\pm$0.12, while in radio-loud the mean is 0.80 with IQR 0.82$\pm$0.13. \par
Finally, we searched for a correlation between the R5007 and the wing velocity. A fast wing might indeed be connected with a reduction of the covering factor in the gas clouds, which translates in a reduction of the equivalent width, and of the flux, in the [O III] lines \citep{Ludwig12}. We found such correlation only in the radio-loud outliers, as shown in Fig.~\ref{wing_r5007}. While in radio-quiet outliers $r =$ -0.4 and p-value = 0.55, among radio-loud the Pearson coefficient is $r =$ -0.8, with a p-value = 4$\times10^{-3}$. The fastest wings, in radio-loud sources, are therefore found in sources with a low R5007.
\section{Discussion}
\subsection{Origin of the radio emission}
\label{subsec:interaction}
A first clear separation between radio-quiet and radio-loud objects is shown in Fig.\ref{radioo3}. The two populations are separated in the plot, both in radio, by construction, and in [O III] luminosity, as confirmed by the K-S test. Nonetheless, as previously mentioned, the [O III] luminosity difference is likely a selection effect. Conversely, the radio emission is so different that it probably has a different origin in the two classes. In the radio-loud sample the emission is likely to be radio jet, together with the radio emission coming from the accretion disk, the corona and a strong starburst component \citep{Caccianiga15}. In radio-quiet sources instead the jet is probably absent or very weak (see Sect.~\ref{subsec:rq}). The radio photons are likely thermal radiation originated via bremmsstrahlung and coming from corona and accretion disk, again with a starburst component and, in some cases, also a faint nonthermal radiation \citep{Giroletti09}. \par
\begin{figure}[t!]
\centering
\includegraphics[width=\hsize]{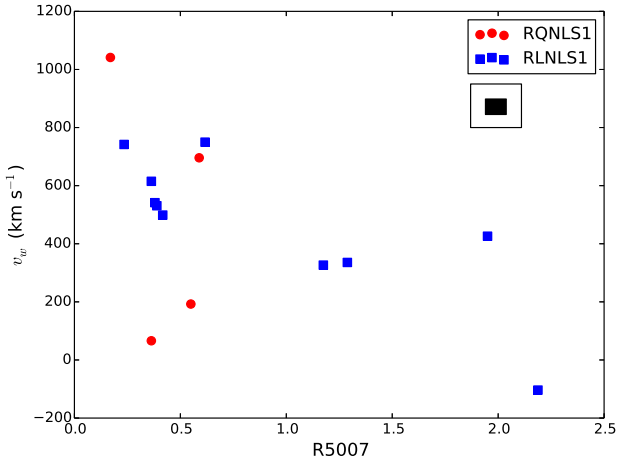} 
\caption{Distribution of R5007 against the wing velocity in outliers sources with a visible wing. RQNLS1s are indicated by red circles, RLNLS1s are indicated by blue squares. Black filled rectangles as in Fig.~\ref{core-wing}.}
\label{wing_r5007}
\end{figure}
There is a larger intrinsic scatter in radio luminosities for radio-loud sources that is proved by the absence of correlation between radio and [O III]. A plausible explanation is an observational effect. Flat-spectrum RLNLS1s are highly variable because of their beamed jet \citep{Foschini15}. This might indeed significantly alter the radio luminosity of the sources, depending on its activity. It is also worth noting that there are few radio-loud sources well overlapped with the RQNLS1s distribution. These might be interpreted as transition objects in which part of the radio emission, particularly in those whose distance is small with respect to the radio-quiet best-fit line, might have the same origin as that of RQNLS1s. Ten of these in particular are distant less than 1$\sigma$ from the best-fit line, and eight of these ten were investigated by \citet{Caccianiga15}. They found an intense star formation, above 20 M$_\odot$/yr, in 6 of these transition objects, and suggested that in such sources star formation might be responsible for a large portion, or even all, of the radio photons. 
\par
\begin{table*}[t!]
\caption{Blue outliers and blue wings in $\gamma$-ray emitters from \citet{Foschini15}, calculated with respect to narrow H$\beta$. }
\label{bogamma}
\centering
\begin{tabular}{l c c c c c }
\hline\hline
Name   & v$_c$ & FWHM$_c$ & v$_w$ & FWHM$_w$ \\
\hline
J0324+3410 & -7.96$\pm$79.04 & 445.59$\pm$116.76 & -123.73$\pm$565.83 & 1201.40$\pm$341.90 \\
J0849+5108 & 264.47$\pm$10.78 & 231.26$\pm$8.98 & -498.77$\pm$74.25 & 703.64$\pm$50.30 \\
J0948+0022 & -770.79$\pm$62.87 & 1438.29$\pm$55.69 & $-$ & $-$ \\
J1102+2239 &  -565.41$\pm$21.56 & 879.90$\pm$11.38 & -749.79$\pm$164.06 & 1319.85$\pm$83.23 \\
J1246+0238 &  39.94$\pm$20.96 & 382.14$\pm$20.36 & -503.49$\pm$136.52 & 1245.11$\pm$70.06 \\
J1505+0326 &  -352.25$\pm$59.28 & 597.88$\pm$60.48 & -326.51$\pm$317.35 & 1180.25$\pm$131.13 \\
J1644+2619 &  -9.76$\pm$26.35 & 145.24$\pm$9.58 & $-$ & $-$ \\
\hline
\end{tabular}
\tablefoot{Columns: (1) name of the source; (2) velocity of the [O III]\seconda\ core component (\kms); (3) FWHM of the core component (\kms); (4) shift of the blue wing with respect to the core component (\kms); (5) FWHM of the blue wing (\kms).}
\end{table*}
\subsection{[O III] lines properties}
The blue wings distribution seems to be roughly the same in the two samples with on average slightly bluer wings in radio-loud sources. This result is expected if the blue wings always originate in outflows, whose velocity is similar in radio-loud and radio-quiet sources. In those sources harboring a relativistic jet, if we admit that an energy transfer occurs between the jet and the NLR medium, the bulk of the gas is accelerated. In this way the two [O III] components, core and wing, are both blueshifted but they mantain the same relative velocity. This acceleration process might also explain the significantly larger ($\sim$2 times larger) velocity IQR in the radio-loud blue wings. \par
Blue outliers are instead more common in the radio-loud than in the radio-quiet sample. Their number is systematically larger regardless of the errors. The simplest explanation for this is that the gas in the NLR of RLNLS1s is often turbulent, possibly because of the interaction between the relativistic jet and the medium. Nonetheless, it must be underlined that the interaction with the NLR does not necessarily occur every time that the jet is present. Some examples are the three RLNLS1s investigated by \citet{Richards15}; none of these are outliers, even if they show alternatively very fast or turbulent blue wings. This interpretation is strengthened by the absence of correlation between radio luminosity and blue outliers in radio-loud sources. While blue outliers are more common among radio-loud sources, apparently a large radio luminosity does not automatically imply a shift in the [O III] core. The age of NLS1s might provide an interesting possibility; if some of these sources have really young jets, the NLR has been interacting with them for a short time and, therefore, the influence over its kinematic might be still negligible. \par

\subsection{Jet/NLR interaction}
Another sign that interactions are effectively ongoing in RLNLS1s is the [O III]/H$\beta$ ratio shown in Fig.\ref{o3hb}. A strong outflow can indeed reduce the equivalent width of the lines by reducing the covering factor of the NLR clouds \citep{Ludwig12}. This decreases the [O III] flux. The H$\beta$ line, coming mostly from the inner part of the BLR, instead remains roughly constant. The ratio should therefore be lower in those sources where the jet interacts with the NLR, forming several fast outflows that decrease the [O III] flux.\par
Our data seem to reveal that the R5007 distribution in both samples is the same (A-D test p-value 0.09). We found that 10 out of 16 radio-loud outliers (63\%), both blue and red, have a R5007 below the average of their sample. Among radio-quiet sources instead two out of four outliers are below the sample average (50\%). If we assume that this difference is real, that is the mechanism reducing the covering factor is not the same in the two samples, this might suggest that the mechanism active in radio-quiet sources is slightly less efficient than the relativistic jet of RLNLS1s. Of course this result must be taken with care, since we are dealing with low numbers. \par
An evident example for the covering factor reduction is J0948+0022. This source was the first $\gamma$-ray NLS1 discovered \citep{Abdo09a}, and it harbors a relativistic beamed jet. Its spectrum barely shows the \seconda\ line and, therefore, the ratio of [O III]/H$\beta$ is only an upper limit ($\lesssim$0.1). Anyway, as shown before, some RLNLS1s do not exhibit any significant [O III] flux and equivalent width reduction. Interestingly we found that all these sources are those showing relatively slow wings in their line profile, while this statement is not true for the two radio-quiet outliers. We indeed tested the correlation between R5007 against the wing velocity, when the latter could be measured. The correlation is present only in RLNLS1s (see Fig.~\ref{wing_r5007}), and this might indicate that in RLNLS1s only sources with a fast outflowing gas show the [O III] covering factor reduction. The different behavior of RQNLS1s might be another hint that the origin of the gas turbulence within the NLR is different. \par
Hence the NLR of these objects is possibly slightly less perturbed than in other radio-loud blue outliers. This might be because of the lack of interaction processes between the jet and the medium. If NLS1s are young sources, it is indeed possible that the [O III] covering factor is not yet affected by the jet moving through the NLR. This possibility however opens a new question: how does the interaction occur? \par
An interesting hypothesis regarding the nature of the interaction can be found in \citet{Morganti15}, and also in the simulations ran by \citet{Wagner11} and \citet{Wagner12}. The jet strongly affects the behavior of the clumpy gas, by following the least resistance path through the clouds and accelerating them in many directions. Around the jet axis a dense and turbulent gas cocoon is formed, that later moves away from the axis and gets dispersed. This increases the turbulence and hence the width of the lines. Since the line of sight in NLS1s is close to the NLR axis, the gas acceleration toward the outer regions appears as a blueshift that should affect all the forbidden lines of the NLR. All these effects should be particularly evident in all the high ionization lines. According to the stratification model \citep{Osterbrock91}, the clouds where they form are the first to interact with the jet. This is particularly true in the [O III] lines case, as they are the strongest high ionization lines in the optical spectrum. \citet{Wagner12} showed that the interaction is active only if the ratio between the jet power and Eddington luminosity of the black hole is high enough ($\log(P_j/L_{Edd}) > -4$). The outliers in our sample whose jet power is known \citep[see][]{Foschini15} always respect this condition. This is in agreement with the perturbed NLR we observe, and it seems to provide an observational confirmation of their results. \par 
Nevertheless, this model still has some issues. Some flat-spectrum RLNLS1s are actually extremely compact, with a typical size below 1 pc. Such small sizes indicate that these sources have not developed radio lobes and the chances for an ongoing interaction seem reduced. We speculate that there is a reasonable mechanism that might allow the jet/NLR feedback in absence of radio lobes. As suggested by \citet{Cavaliere02} for blazars, the jet engine might be a combination of the Blandford-Znajek (BZ) and the Blandford-Payne (BP) mechanisms \citep{mecBZ, BlandfordPayne}. Also in the case of NLS1s, the BZ alone is not enough to explain the observed jet power \citep{Foschini11b}. Therefore, the BP is likely providing the required additional power. Since the BP is essentially a centrifugally driven wind that becomes more efficient when the accretion luminosity is high, it can significantly contribute to the acceleration of the gas in the NLR, even in absence of radio lobes. The wind can be generated only if the poloidal component of the magnetic field forms an angle lower than 60$^\circ$ with the disk plane. If this condition is not matched, the jet is not launched and the source might appear as radio quiet. Despite this, the disk is still able to produce strong winds, that in the most luminous sources can generate outliers.  \par
In our radio-loud sample there are seven $\gamma$-ray emitters \citep{Foschini15, Dammando15}, and four of these sources are outliers (three blue, one red). These results are shown in Tab.~\ref{bogamma}. In two of the remaining sources, J0324+3410 and J1246+0238, the [O III] lines are not particularly shifted, but their blue wings have a high FWHM, which again indicates a very strong internal turbulence. Therefore even if the numbers are still low, more than half of the $\gamma$-ray emitters are also outliers, and this might suggest a connection between these two phenomena. \par

\subsection{Radio quiet versus radio loud}
\label{subsec:rq}
All our findings seem to point out that a relativistic jet has a strong influence on the NLR. Radio-loud sources, and particularly the $\gamma$-ray emitters, indeed show a strongly disturbed NLR kinematics. In radio-quiet sources the NLR is also perturbed, but the number of blue outliers among these sources is significantly lower. We speculate that the BP mechanism is a possible way to account for the differences we found between radio-quiet and radio-loud NLS1s. In NLS1s, without the decisive contribution of the BP, the jet is not launched. Nevertheless this mechanism can actually provide a contribution to the collimation only if the critical angle criterium is met. In RQNLS1s the poloidal component of the magnetic field might have a high inclination and, therefore, the disk can only form a jet base \citep{Falcke99}. Since it is a pressure driven structure, the jet base cannot accelerate the plasma particles to relativistic speed, but only to moderate velocities. Therefore a nonrelativistic wind is present, which is only powerful enough in the most luminous sources to affect the kinematic of the whole NLR. In general, instead, this outflowing wind might be the region where blue wings are formed. This mechanism is in agreement with our finding that only sources with a radio emission above 10$^{39}$ \ergs\ can be outliers. This high radio luminosity might indicate in RQNLS1s the presence of a strong jet base, which might be more effective in perturbing the gas dynamics. The same mechanism can explain the presence of the wings in all Seyfert galaxies: both Seyfert 1 and 2 indeed show this feature in their [O III] lines. \par
In RLNLS1s, instead, the condition on the critical angle is typically met. Therefore the poloidal component of the magnetic field accelerates the particles, and at the same time it forms the relativistic jet and surrounding wind. The accretion disk of NLS1s is luminous enough to be in the radiation pressure dominated regime in which radiative instability occurs \citep{Moderski96, Ghosh97, Czerny09, Wu09b, Foschini11b}. This instability might translate into a change in angle, and then in an intermittent BP contribution to the jet. The strong winds are instead always present and, because they are directed along the jet axis, they influence the NLR kinematics even when the jet is not active, providing the observed large number of outliers. 

\subsection{Implications for the parent population}
Our results likely provide an important contribution to the search of flat-spectrum RLNLS1s parent population. The hypothesis of \citet{Foschini11} was that in RQNLS1s the jet is highly collimated and, since its energy is not dissipated, it is invisible for present-day observatories. If this is true, the dichotomy between radio-quiet and radio-loud NLS1s might only be due to an orientation effect. This might explain why the NLR of RQNLS1s appears to be less perturbed. One explanation of why the NLR of RQNLS1s appears to be less perturbed is that in RQNLS1s low power jets are present, that do not dissipate their energy for a long time. As previously mentioned, some traces of nonthermal radio emission are actually present in some RQNLS1s. In particular, \citet{Giroletti09} found this emission in one of our sources, J1203+4431 (NGC 4051). Anyway its flux is very weak and it can be explained simply with the jet base model without invoking a fully developed relativistic jet. \par
Therefore, since RQNLS1s do not seem to harbor a fully relativistic jet, they may not be considered part of the parent population of radio-loud NLS1s. Of course there might be a few exceptions. Sources that, even though they are radio-quiet and actually show a fully developed radio jet, do actually exist (i.e. Mrk 1239, \citealp{Doi15}). Nevertheless this might happen because, as shown in \citet{Ho01}, radio loudness can be a slightly misleading parameter, since it is strongly affected by the way of measuring both the radio and optical flux. It is also possible that the source activity was relatively low at the time of observation, while the jets were formed in the past, when the BP was active. Therefore radio quietness for some sources might be a simply temporary condition.

\section{Summary}
In this work we investigated the nature of the parent population of flat-spectrum RLNLS1s by means of the [O III] line properties in two samples of NLS1s: one radio loud and one radio quiet. Such a study can provide important information on the NLR kinematics and the interactions between a relativistic jet and its environment. It can finally help us to understand whether a fully developed relativistic jet is also present in RQNLS1s and, therefore, whether these objects belong to the parent population. \par
We decomposed each [O III] line into two components, one to represent the core and another to represent the wing. The first is dominated by the bulge gravitational potential, while the second is likely generated by outflows coming from the inner NLR. The wings have roughly the same relative velocity in both our samples, and it appears that faster outflows are typically associated with a large internal turbulence. The core component is also affected by the disturbed kinematics of the gas, and this is particularly evident in radio-loud sources that harbor a relativistic jet. A large portion of these are in fact blue outliers. \par
We interpret these results as a hint of an ongoing interaction between the NLR and the relativistic jets in RLNLS1s. The jet likely accelerates the gas in the NLR, generating a large number of blue and red outliers. These appear to be more common among RL than RQNLS1s. Another effect of the jet is the reduction of equivalent width of high ionization lines in blue outliers, which in some cases translates to a reduction of the [O III]/H$\beta$ ratio. Moreover, we also found that there might be a connection between blue outliers and $\gamma$-ray emission. Out of seven $\gamma$-ray emitting NLS1s included in our sample, we indeed found four outliers. \par
Since not all RLNLS1s developed radio lobes to interact with the medium, we speculate that the interaction might also occur in a different way. The BP mechanism might provide a valid explanation both for the jet/NLR interaction in RLNLS1s, and also for the absence of a fully developed relativistic jet in RQNLS1s. The BP requires a critical angle between the poloidal component of the magnetic field and the disk surface to launch the jet. When this condition is met the source appears as radio loud, and otherwise it appears as radio quiet. Nevertheless, as the central engine is a continuously varying region, a radio-loud source can go through different states of activity, switching off the BP mechanism and producing an intermittent jet. This mechanism can produce strong winds even when the jet is off. In this way, the NLR kinematics can be perturbed even if radio lobes are not present. Finally, it can also account for the outflows where blue wings are originated. In few cases, these outflows are powerful enough to produce blue outliers even without a jet. \par
If this hypothesis is correct, RQNLS1s do not harbor a fully developed relativistic jet, but only a jet base that can account for the faint nonthermal radio emission observed in these sources. Hence they might not be included in the parent population of flat-spectrum RLNLS1s. Future studies are needed to deeply investigate the innermost region of RQNLS1s, to confirm the absence of the relativistic jet. For this purpose, new generation instruments as JVLA or SKA can help to deeply investigate the matter. \par

\begin{acknowledgements}
We thank our referee David Rosario for helpful suggestions that greatly improved the quality of the paper. This paper is based on observations collected with the 1.22m \textit{Galileo} telescope of the Asiago Astrophysical Observatory, operated by the Department of Physics and Astronomy "G. Galilei" of the University of Padova, and with the Italian Telescopio Nazionale Galileo (TNG) operated on the island of La Palma by the Fundaci\'on Galileo Galilei of the INAF (Istituto Nazionale di Astrofisica) at the Spanish Observatorio del Roque de los Muchachos of the Instituto de Astrofisica de Canarias. This research has made use of the NASA/IPAC Extragalactic Database (NED), which is operated by the Jet Propulsion Laboratory, California Institute of Technology, under contract with the National Aeronautics and Space Administration. Funding for the Sloan Digital Sky Survey has been provided by the Alfred P. Sloan Foundation and the U.S. Department of Energy Office of Science. The SDSS web site is \texttt{http://www.sdss.org}. SDSS-III is managed by the Astrophysical Research Consortium for the Participating Institutions of the SDSS-III Collaboration including the University of Arizona, the Brazilian Participation Group, Brookhaven National Laboratory, Carnegie Mellon University, University of Florida, the French Participation Group, the German Participation Group, Harvard University, the Instituto de Astrofisica de Canarias, the Michigan State/Notre Dame/JINA Participation Group, Johns Hopkins University, Lawrence Berkeley National Laboratory, Max Planck Institute for Astrophysics, Max Planck Institute for Extraterrestrial Physics, New Mexico State University, University of Portsmouth, Princeton University, the Spanish Participation Group, University of Tokyo, University of Utah, Vanderbilt University, University of Virginia, University of Washington, and Yale University. 
\end{acknowledgements}
\bibliographystyle{aa}
\bibliography{./biblio}

\begin{table*}
\caption{Summary of the sources intrinsic properties in RQNLS1s. }
\label{tab:summary1}
\centering
\scalebox{0.9}{
\footnotesize
\begin{tabular}{l c c c c c c c} 
\hline\hline
Short Name & R.A. & Dec. & z & $\log$M$_{BH}$ & Edd & $\log{L}$(H$\beta$) & $\log{L}_{rad}$ \\
\hline\hline
J0306+0003 & 03h06m39.58s & +00d03m43.2 & 0.107 & 7.18 & 0.12 & 41.91 & 39.12 \\
J0632+6340 & 06h32m47.16s & +63d40m52.1 & 0.013 & 6.48 & 0.02 & 40.52 & 37.78 \\
J0736+3926 & 07h36m23.13s & +39d26m17.7 & 0.118 & 7.69 & 0.17 & 42.55 & 39.26 \\
J0751+2914 & 07h51m01.42s & +29d14m19.1 & 0.121 & 7.38 & 0.17 & 42.23 & 38.91 \\
J0752+2617 & 07h52m45.60s & +26d17m35.7 & 0.082 & 7.16 & 0.09 & 41.78 & 38.49 \\
J0754+3920 & 07h54m00.05s & +39d20m29.1 & 0.096 & 8.04 & 0.20 & 42.93 & 39.58 \\
J0804+3853 & 08h04m09.24s & +38d53m48.8 & 0.211 & 7.72 & 0.14 & 42.48 & 39.71 \\
J0818+3834 & 08h18m49.26s & +38d34m16.1 & 0.160 & 6.98 & 0.12 & 41.71 & 39.31 \\
J0836+1554 & 08h36m15.37s & +15d54m09.8 & 0.206 & 7.34 & 0.09 & 41.95 & 39.29 \\
J0913+3658 & 09h13m13.72s & +36d58m17.2 & 0.107 & 7.01 & 0.06 & 41.49 & 38.54 \\
J0925+5217 & 09h25m12.87s & +52d17m10.5 & 0.035 & 7.51 & 0.10 & 42.13 & 38.31 \\
J0926+1244 & 09h26m03.25s & +12d44m04.1 & 0.029 & 7.31 & 0.04 & 41.56 & 38.30 \\
J0936$-$0026 & 09h36m09.13s & $-$00d26m39.7 & 0.141 & 6.94 & 0.07 & 41.49 & 38.72 \\
J0948+5029 & 09h48m42.67s & +50d29m31.4 & 0.056 & 7.02 & 0.07 & 41.52 & 38.79 \\
J0957+2433 & 09h57m07.16s & +24d33m16.1 & 0.082 & 6.98 & 0.09 & 41.59 & 38.56 \\
J0958+5602 & 09h58m33.94s & +56d02m24.4 & 0.216 & 7.42 & 0.05 & 41.74 & 39.18 \\
J1016+4210 & 10h16m45.11s & +42d10m25.5 & 0.055 & 7.14 & 0.08 & 41.71 & 38.18 \\
J1022+2022 & 10h22m58.20s & +20d22m37.9 & 0.130 & 7.41 & 0.03 & 41.62 & 38.70 \\
J1025+5140 & 10h25m31.28s & +51d40m34.9 & 0.045 & 7.16 & 0.07 & 41.64 & 37.85 \\
J1036+4125 & 10h36m04.66s & +41d25m17.8 & 0.120 & 7.05 & 0.05 & 41.46 & 38.79 \\
J1050+1132 & 10h50m07.75s & +11d32m28.6 & 0.133 & 7.81 & 0.15 & 42.58 & 39.00 \\
J1103+0834 & 11h03m33.00s & +08d34m49.0 & 0.163 & 7.10 & 0.06 & 41.55 & 39.24 \\
J1112+4541 & 11h12m39.56s & +45d41m41.3 & 0.136 & 6.94 & 0.19 & 41.88 & 39.01 \\
J1120+0633 & 11h20m14.85s & +06d33m41.1 & 0.316 & 7.36 & 0.22 & 42.33 & 39.90 \\
J1121+5351 & 11h21m08.59s & +53d51m21.1 & 0.103 & 7.64 & 0.13 & 42.37 & 39.03 \\
J1128+1023 & 11h28m13.02s & +10d23m08.3 & 0.050 & 6.91 & 0.07 & 41.43 & 37.98 \\
J1136+3432 & 11h36m55.95s & +34d32m37.0 & 0.192 & 7.16 & 0.23 & 42.16 & 39.38 \\
J1149+0448 & 11h49m54.98s & +04d48m12.8 & 0.270 & 7.94 & 0.10 & 42.54 & 39.74 \\
J1155+1507 & 11h55m23.74s & +15d07m56.9 & 0.287 & 7.79 & 0.16 & 42.61 & 39.86 \\
J1203+4431 & 12h03m09.69s & +44d31m52.5 & 0.002 & 6.30 & 0.01 & 39.97 & 36.18 \\
J1207$-$0219 & 12h07m00.30s & $-$02d19m27.1 & 0.308 & 7.45 & 0.13 & 42.20 & 39.53 \\
J1209+3217 & 12h09m45.20s & +32d17m01.1 & 0.144 & 7.48 & 0.10 & 42.13 & 39.26 \\
J1215+5442 & 12h15m49.44s & +54d42m24.0 & 0.150 & 7.51 & 0.10 & 42.13 & 39.26 \\
J1218+1834 & 12h18m30.84s & +18d34m58.2 & 0.197 & 7.77 & 0.11 & 42.43 & 39.04 \\
J1218+2948 & 12h18m26.48s & +29d48m46.2 & 0.013 & 6.43 & 0.05 & 40.90 & 38.31 \\
J1242+3317 & 12h42m10.61s & +33d17m02.6 & 0.044 & 7.06 & 0.06 & 41.52 & 38.57 \\
J1246+0222 & 12h46m35.25s & +02d22m08.8 & 0.048 & 7.09 & 0.05 & 41.49 & 38.17 \\
J1311+0648 & 13h11m56.15s & +06d48m58.3 & 0.128 & 7.11 & 0.11 & 41.82 & 39.07 \\
J1315+4325 & 13h15m10.07s & +43d25m47.0 & 0.086 & 6.85 & 0.08 & 41.41 & 38.70 \\
J1320+2108 & 13h20m46.67s & +21d08m46.4 & 0.090 & 7.06 & 0.05 & 41.44 & 38.65 \\
J1322+0809 & 13h22m55.43s & +08d09m41.6 & 0.050 & 7.20 & 0.06 & 41.61 & 38.74 \\
J1331+0131 & 13h31m38.03s & +01d31m51.6 & 0.080 & 6.82 & 0.05 & 41.20 & 38.83 \\
J1332+3127 & 13h32m05.28s & +31d27m36.4 & 0.090 & 7.25 & 0.11 & 41.96 & 39.25 \\
J1337+2423 & 13h37m18.72s & +24d23m03.4 & 0.108 & 8.04 & 0.22 & 42.97 & 39.90 \\
J1342+0505 & 13h42m06.56s & +05d05m23.8 & 0.266 & 7.80 & 0.16 & 42.61 & 40.07 \\
J1342+4642 & 13h42m43.57s & +46d42m24.0 & 0.086 & 6.90 & 0.07 & 41.45 & 38.52 \\
J1355+5612 & 13h55m16.56s & +56d12m44.6 & 0.122 & 7.14 & 0.20 & 42.08 & 39.53 \\
J1358+2511 & 13h58m52.00s & +25d11m40.2 & 0.089 & 7.29 & 0.06 & 41.74 & 38.88 \\
J1402+1720 & 14h02m59.03s & +17d20m56.0 & 0.060 & 6.60 & 0.07 & 41.16 & 38.44 \\
J1402+2159 & 14h02m34.44s & +21d59m51.5 & 0.066 & 7.15 & 0.11 & 41.84 & 38.11 \\
J1406+2223 & 14h06m21.89s & +22d23m46.5 & 0.098 & 7.60 & 0.09 & 42.19 & 38.90 \\
J1408+2409 & 14h08m27.82s & +24d09m24.6 & 0.130 & 7.15 & 0.16 & 41.99 & 39.28 \\
J1439+3923 & 14h39m52.91s & +39d23m58.9 & 0.112 & 7.23 & 0.05 & 41.56 & 38.88 \\
J1440+6156 & 14h40m12.74s & +61d56m33.0 & 0.275 & 7.93 & 0.21 & 42.84 & 39.98 \\
J1441+1604 & 14h41m56.56s & +16d04m21.1 & 0.113 & 7.07 & 0.07 & 41.61 & 38.89 \\
J1442+2623 & 14h42m40.79s & +26d23m32.5 & 0.107 & 7.10 & 1.22 & 41.66 & 39.15 \\
J1444+1536 & 14h44m31.62s & +15d36m43.2 & 0.050 & 6.89 & 0.05 & 41.27 & 38.54 \\
J1448+3559 & 14h48m25.09s & +35d59m46.6 & 0.113 & 7.57 & 0.07 & 42.05 & 38.84 \\
J1451+2709 & 14h51m08.76s & +27d09m26.9 & 0.065 & 7.32 & 0.15 & 42.13 & 38.72 \\
J1536+5433 & 15h36m38.39s & +54d33m33.2 & 0.039 & 7.34 & 0.09 & 41.96 & 37.79 \\
J1537+4942 & 15h37m32.62s & +49d42m47.7 & 0.280 & 7.36 & 0.12 & 42.07 & 39.75 \\
J1555+1911 & 15h55m07.92s & +19d11m32.4 & 0.035 & 6.50 & 0.03 & 40.80 & 37.85 \\
J1559+3501 & 15h59m09.63s & +35d01m47.5 & 0.031 & 6.86 & 0.06 & 41.33 & 38.02 \\
J1605+3239 & 16h05m08.87s & +32d39m21.4 & 0.091 & 7.13 & 0.02 & 41.15 & 38.49 \\
J1627+4736 & 16h27m50.54s & +47d36m23.5 & 0.262 & 8.03 & 0.12 & 42.73 & 40.08 \\
J2140+0025 & 21h40m54.55s & +00d25m38.1 & 0.084 & 7.16 & 0.12 & 41.88 & 38.27 \\
J2219+1207 & 22h19m18.53s & +12d07m53.1 & 0.081 & 6.92 & 0.06 & 41.38 & 38.54 \\
J2254+0046 & 22h54m52.22s & +00d46m31.3 & 0.091 & 7.06 & 0.06 & 41.51 & 38.28 \\
\hline
\end{tabular}
}
\tablefoot{Columns: (1) short name; (2) right ascension; (3) declination; (4) redshift; (5) logarithm of the black hole mass; (6) Eddington ratio; (7) logarithm of the H$\beta$ luminosity; and (8) logarithm of the radio luminosity at 1.4 GHz.}
\end{table*}
\clearpage
\begin{table*}
\caption{Summary of the sources intrinsic properties in RLNLS1s. }
\label{tab:summary2}
\centering
\scalebox{0.9}{
\footnotesize
\begin{tabular}{l c c c c c c c} 
\hline\hline
Short Name & R.A. & Dec. & z & $\log$M$_{BH}$ & Edd & $\log{L}$(H$\beta$) & $\log{L}_{rad}$ \\
\hline\hline
J0138+1321 & 01h38m59.33s & +13d21m08.2 & 0.243 & 7.49 & 0.09 & 42.11 & 40.27 \\
J0146$-$0040 & 01h46m44.82s & -00d40m43.1 & 0.083 & 7.32 & 0.07 & 41.79 & 39.03 \\
J0251$-$0702 & 02h51m05.28s & -07d02m30.1 & 0.327 & 7.54 & 0.12 & 42.25 & 40.47 \\
J0324+3410 & 03h24m41.16s & +34d10m45.8 & 0.061 & 7.67 & 0.03 & 41.73 & 40.89 \\
J0706+3901 & 07h06m25.15s & +39d01m51.6 & 0.086 & 7.04 & 0.04 & 41.34 & 39.16 \\
J0713+3820 & 07h13m40.29s & +38d20m40.1 & 0.123 & 8.20 & 0.17 & 43.02 & 39.76 \\
J0804+3853 & 08h04m09.24s & +38d53m48.8 & 0.211 & 8.00 & 0.09 & 42.57 & 39.71 \\
J0806+7248 & 08h06m38.96s & +72d48m20.4 & 0.098 & 6.94 & 0.08 & 41.52 & 40.23 \\
J0814+5609 & 08h14m32.11s & +56d09m56.6 & 0.509 & 8.44 & 0.11 & 43.09 & 41.99 \\
J0849+5108 & 08h49m57.97s & +51d08m29.0 & 0.584 & 7.37 & 0.13 & 42.12 & 42.83 \\
J0850+4626 & 08h50m01.17s & +46d26m00.5 & 0.524 & 7.83 & 0.09 & 42.41 & 41.50 \\
J0902+0443 & 09h02m27.16s & +04d43m09.5 & 0.532 & 7.70 & 0.12 & 42.39 & 42.38 \\
J0937+3615 & 09h37m09.02s & +36d15m37.1 & 0.179 & 7.58 & 0.05 & 41.96 & 39.66 \\
J0948+0022 & 09h48m57.31s & +00d22m25.4 & 0.585 & 7.82 & 0.15 & 42.62 & 42.32 \\
J0952$-$0136 & 09h52m19.17s & -01d36m44.1 & 0.020 & 7.13 & 0.05 & 41.37 & 38.87 \\
J0953+2836 & 09h53m17.09s & +28d36m01.5 & 0.658 & 8.51 & 0.04 & 42.72 & 42.07 \\
J1031+4234 & 10h31m23.73s & +42d34m39.3 & 0.379 & 8.46 & 0.02 & 42.43 & 41.06 \\
J1034+3938 & 10h34m38.60s & +39d38m27.8 & 0.043 & 6.03 & 0.17 & 40.97 & 39.20 \\
J1037+0036 & 10h37m27.45s & +00d36m35.6 & 0.595 & 7.48 & 0.14 & 42.25 & 41.75 \\
J1038+4227 & 10h38m59.58s & +42d27m42.2 & 0.220 & 7.88 & 0.08 & 42.42 & 40.15 \\
J1047+4725 & 10h47m32.68s & +47d25m32.0 & 0.798 & 8.39 & 0.08 & 42.90 & 43.49 \\
J1048+2222 & 10h48m16.58s & +22d22m39.0 & 0.330 & 7.53 & 0.11 & 42.20 & 39.77 \\
J1102+2239 & 11h02m23.39s & +22d39m20.7 & 0.453 & 8.17 & 0.06 & 42.59 & 40.34 \\
J1110+3653 & 11h10m05.03s & +36d53m36.3 & 0.630 & 7.09 & 0.23 & 42.10 & 41.64 \\
J1133+0432 & 11h33m20.91s & +04d32m55.1 & 0.248 & 7.28 & 0.12 & 42.04 & 40.12 \\
J1138+3653 & 11h38m24.54s & +36d53m27.1 & 0.356 & 7.61 & 0.09 & 42.20 & 40.88 \\
J1146+3236 & 11h46m54.28s & +32d36m52.3 & 0.465 & 8.18 & 0.07 & 42.60 & 41.22 \\
J1200$-$0046 & 12h00m14.08s & -00d46m38.7 & 0.210 & 7.81 & 0.08 & 42.33 & 40.85 \\
J1227+3214 & 12h27m49.14s & +32d14m58.9 & 0.137 & 6.84 & 0.16 & 41.71 & 39.66 \\
J1238+3942 & 12h38m52.12s & +39d42m27.8 & 0.623 & 6.82 & 0.44 & 42.10 & 41.37 \\
J1246+0238 & 12h46m34.65s & +02d38m09.0 & 0.363 & 7.94 & 0.06 & 42.35 & 41.37 \\
J1302+1624 & 13h02m58.77s & +16d24m27.6 & 0.067 & 7.36 & 0.12 & 42.08 & 39.45 \\
J1305+5116 & 13h05m22.74s & +51d16m40.2 & 0.788 & 8.20 & 0.67 & 43.58 & 42.53 \\
J1333+4141 & 13h33m45.47s & +41d41m27.7 & 0.225 & 7.92 & 0.05 & 42.23 & 39.71 \\
J1337+6005 & 13h37m24.32s & +60d05m41.7 & 0.234 & 6.67 & 0.28 & 41.78 & 40.02 \\
J1346+3121 & 13h46m34.97s & +31d21m33.7 & 0.246 & 7.20 & 0.10 & 41.86 & 39.50 \\
J1358+2658 & 13h58m45.38s & +26d58m08.5 & 0.331 & 7.84 & 0.12 & 42.52 & 39.77 \\
J1409+5656 & 14h09m14.35s & +56d56m25.7 & 0.239 & 7.78 & 0.03 & 41.89 & 39.83 \\
J1432+3014 & 14h32m44.91s & +30d14m35.3 & 0.355 & 7.48 & 0.21 & 42.43 & 41.47 \\
J1435+3131 & 14h35m09.49s & +31d31m47.8 & 0.502 & 7.56 & 0.12 & 42.28 & 41.73 \\
J1443+4725 & 14h43m18.56s & +47d25m56.7 & 0.706 & 7.93 & 0.12 & 42.62 & 42.71 \\
J1450+5919 & 14h50m41.93s & +59d19m36.9 & 0.202 & 7.04 & 0.11 & 41.74 & 39.73 \\
J1505+0326 & 15h05m06.47s & +03d26m30.8 & 0.409 & 7.26 & 0.06 & 41.70 & 42.48 \\
J1507+4453 & 15h07m40.92s & +44d53m31.5 & 0.314 & 7.45 & 0.16 & 42.28 & 40.01 \\
J1358+2658 & 13h58m45.38s & +26d58m08.4 & 0.331 & 7.62 & 0.19 & 42.52 & 39.77 \\
J1548+3511 & 15h48m17.92s & +35d11m28.0 & 0.479 & 7.96 & 0.13 & 42.69 & 42.23 \\
J1608+0708 & 16h08m31.56s & +07d08m18.2 & 0.153 & 7.62 & 0.02 & 41.49 & 39.17 \\
J1612+4219 & 16h12m59.83s & +42d19m40.3 & 0.234 & 6.68 & 0.41 & 41.97 & 39.91 \\
J1629+4007 & 16h29m01.30s & +40d07m59.9 & 0.272 & 7.83 & 0.13 & 42.55 & 40.59 \\
J1633+4718 & 16h33m23.58s & +47d18m58.9 & 0.116 & 6.91 & 0.11 & 41.61 & 40.49 \\
J1634+4809 & 16h34m01.94s & +48d09m40.2 & 0.495 & 7.86 & 0.08 & 42.38 & 40.99 \\
J1644+2619 & 16h44m42.53s & +26d19m13.2 & 0.145 & 6.95 & 0.11 & 41.68 & 40.84 \\
J1703+4540 & 17h03m30.38s & +45d40m47.1 & 0.060 & 7.73 & 0.01 & 41.44 & 40.14 \\
J1709+2348 & 17h09m07.80s & +23d48m37.6 & 0.254 & 7.57 & 0.06 & 42.03 & 39.64 \\
J1713+3523 & 17h13m04.46s & +35d23m33.5 & 0.084 & 6.69 & 0.06 & 41.20 & 39.44 \\
J1722+5654 & 17h22m06.03s & +56d54m51.6 & 0.426 & 7.89 & 0.09 & 42.44 & 41.53 \\
J2314+2243 & 23h14m55.89s & +22d43m25.7 & 0.169 & 8.00 & 0.17 & 42.82 & 40.29 \\
\hline
\end{tabular}
}
\tablefoot{Columns as in Table~\ref{tab:summary1}.}
\end{table*}
\clearpage

\begin{table*}
\caption{Summary of the [O III] line properties for RQNLS1s. }
\label{tab:summary3}
\centering
\scalebox{0.9}{
\begin{tabular}{l c l l l l l l}
\hline\hline
Short Name & $\log{L}_{[O III]}$ & $\lambda_c$ & v$_c$ & FWHM$_c$ & $\lambda_w$ & v$_w$ & FWHM$_w$ \\
\hline\hline
J0306+0003 & 41.69 & 5007.32$\pm$0.08 & 28.56$\pm$4.79 & 382.14$\pm$4.79 & 5003.32$\pm$0.43 & -239.25$\pm$28.74 & 926.43$\pm$13.77 \\
J0632+6340* & 40.28 & 5007.03$\pm$0.22 & 11.20$\pm$13.17 & 547.12$\pm$4.19 & 5002.18$\pm$3.92 & -290.47$\pm$242.50 & 999.76$\pm$115.56 \\
J0736+3926 & 42.12 & 5008.11$\pm$0.09 & 75.86$\pm$5.39 & 423.03$\pm$2.40 & 5002.51$\pm$0.26 & -335.12$\pm$19.76 & 633.13$\pm$5.39 \\
J0751+2914 & 41.68 & 5005.53$\pm$0.09 & -78.62$\pm$5.39 & 283.43$\pm$4.79 & 4999.94$\pm$0.27 & -334.93$\pm$19.76 & 987.06$\pm$15.57 \\
J0752+2617 & 41.00 & 5006.13$\pm$0.06 & -42.69$\pm$3.59 & 227.02$\pm$2.40 & 5001.64$\pm$0.31 & -268.66$\pm$20.36 & 720.56$\pm$8.98 \\
J0754+3920 & 42.67 & 5005.40$\pm$0.07 & -86.40$\pm$4.19 & 644.41$\pm$2.99 & 4993.81$\pm$0.40 & -694.39$\pm$26.35 & 468.15$\pm$17.96 \\
J0804+3853 & 42.19 & 5007.01$\pm$0.04 & 10.00$\pm$2.40 & 157.93$\pm$1.20 & 5000.52$\pm$0.15 & -388.22$\pm$9.58 & 760.04$\pm$3.59 \\
J0818+3834 & 41.49 & 5006.04$\pm$0.08 & -48.08$\pm$4.79 & 289.07$\pm$4.19 & 5000.76$\pm$0.61 & -316.14$\pm$39.52 & 820.67$\pm$14.97 \\
J0836+1554 & 41.47 & 5006.07$\pm$0.16 & -46.28$\pm$9.58 & 353.93$\pm$9.58 & 4999.59$\pm$0.59 & -388.32$\pm$43.11 & 795.29$\pm$11.38 \\
J0913+3658 & 41.43 & 5005.90$\pm$0.08 & -56.46$\pm$4.79 & 329.96$\pm$5.39 & 5000.78$\pm$0.32 & -306.62$\pm$22.15 & 799.52$\pm$7.78 \\
J0925+5217 & 41.99 & 5007.78$\pm$0.53 & 56.10$\pm$31.73 & 215.74$\pm$1.80 & $-$ & $-$ & $-$ \\
J0926+1244* & 41.14 & 5006.32$\pm$0.61 & -31.32$\pm$36.52 & 337.01$\pm$31.73 & 5005.00$\pm$0.96 & -79.02$\pm$92.21 & 846.06$\pm$32.93 \\
J0936$-$0026 & 41.21 & 5006.35$\pm$0.16 & -29.52$\pm$9.58 & 258.05$\pm$16.77 & 5001.99$\pm$0.32 & -261.29$\pm$26.94 & 1050.52$\pm$10.78 \\
J0948+5029 & 41.48 & 5006.74$\pm$0.05 & -6.17$\pm$2.99 & 337.01$\pm$1.80 & 5001.76$\pm$0.17 & -298.60$\pm$11.38 & 868.62$\pm$5.39 \\
J0957+2433 & 41.36 & 5005.64$\pm$0.06 & -72.03$\pm$3.59 & 327.14$\pm$2.40 & 5001.11$\pm$0.27 & -271.66$\pm$17.96 & 958.86$\pm$9.58 \\
J0958+5602 & 41.74 & 5007.96$\pm$0.07 & 66.88$\pm$4.19 & 329.96$\pm$6.59 & 5005.26$\pm$0.71 & -161.79$\pm$44.91 & 1026.55$\pm$62.87 \\
J1016+4210 & 40.97 & 5005.04$\pm$0.06 & -107.96$\pm$3.59 & 201.64$\pm$2.99 & 5002.30$\pm$0.59 & -164.08$\pm$37.72 & 813.62$\pm$28.14 \\
J1022+2022 & 41.27 & 5006.25$\pm$0.19 & -35.51$\pm$11.38 & 245.36$\pm$20.36 & 5004.56$\pm$0.80 & -101.08$\pm$57.48 & 592.24$\pm$55.69 \\
J1025+5140 & 40.88 & 5006.28$\pm$0.10 & -33.71$\pm$5.99 & 236.90$\pm$6.59 & 5003.10$\pm$0.85 & -190.64$\pm$55.09 & 760.04$\pm$56.88 \\
J1036+4125 & 41.28 & 5006.52$\pm$0.06 & -19.34$\pm$3.59 & 331.37$\pm$4.19 & 5002.57$\pm$0.75 & -236.81$\pm$46.70 & 1033.60$\pm$22.15 \\
J1050+1132 & 42.14 & 5005.46$\pm$0.10 & -82.81$\pm$5.99 & 353.93$\pm$4.19 & 5000.81$\pm$0.27 & -278.44$\pm$19.76 & 717.74$\pm$8.38 \\
J1103+0834 & 41.49 & 5005.64$\pm$0.15 & -72.03$\pm$8.98 & 358.16$\pm$13.77 & 5000.82$\pm$0.52 & -289.04$\pm$38.32 & 886.95$\pm$16.17 \\
J1112+4541 & 41.62 & 5004.17$\pm$0.06 & -160.05$\pm$3.59 & 274.97$\pm$1.80 & 5000.96$\pm$0.14 & -192.49$\pm$9.58 & 981.42$\pm$5.99 \\
J1120+0633 & 42.44 & 5006.81$\pm$0.04 & -1.98$\pm$2.40 & 190.36$\pm$1.20 & 5006.32$\pm$0.07 & -29.64$\pm$4.19 & 678.25$\pm$4.19 \\
J1121+5351 & 42.01 & 5007.09$\pm$0.06 & 14.79$\pm$3.59 & 305.99$\pm$2.99 & 5003.61$\pm$0.15 & -208.07$\pm$10.78 & 734.66$\pm$4.79 \\
J1128+1023 & 41.24 & 5006.97$\pm$0.04 & 7.60$\pm$2.40 & 143.83$\pm$1.20 & 5001.36$\pm$0.08 & -335.46$\pm$5.39 & 679.66$\pm$2.40 \\
J1136+3432 & 42.07 & 5006.01$\pm$0.04 & -49.88$\pm$2.40 & 242.54$\pm$1.20 & 5004.29$\pm$0.09 & -102.62$\pm$6.59 & 781.19$\pm$5.39 \\
J1149+0448* & 41.77 & 5018.45$\pm$0.17 & 694.99$\pm$10.18 & 166.39$\pm$70.06 & 5001.02$\pm$0.24 & -1041.24$\pm$22.75 & 1191.53$\pm$5.39 \\
J1155+1507 & 42.31 & 5007.11$\pm$0.08 & 15.99$\pm$4.79 & 180.49$\pm$4.79 & 5000.46$\pm$0.14 & -398.26$\pm$11.38 & 943.35$\pm$4.79 \\
J1203+4431 & 39.94 & 5006.94$\pm$0.45 & 5.81$\pm$26.94 & 318.68$\pm$5.99 & 5002.02$\pm$0.91 & -294.61$\pm$59.88 & 606.34$\pm$81.43 \\
J1209+3217* & 41.74 & 5006.17$\pm$0.19 & -40.30$\pm$11.38 & 410.34$\pm$6.59 & 5002.46$\pm$0.24 & -221.92$\pm$23.95 & 1405.86$\pm$12.57 \\
J1207$-$0219 & 42.02 & 5006.13$\pm$0.10 & -42.69$\pm$5.99 & 279.20$\pm$5.39 & 4999.33$\pm$0.21 & -407.22$\pm$16.77 & 1125.25$\pm$7.19 \\
J1215+5442 & 41.78 & 5005.69$\pm$0.16 & -69.04$\pm$9.58 & 384.96$\pm$7.19 & 4997.34$\pm$0.58 & -500.47$\pm$42.51 & 661.33$\pm$15.57 \\
J1218+1834 & 41.97 & 5006.93$\pm$0.07 & 5.21$\pm$4.19 & 320.09$\pm$7.78 & 5003.66$\pm$0.33 & -195.68$\pm$22.15 & 853.11$\pm$11.38 \\
J1218+2948 & 41.26 & 5007.70$\pm$0.38 & 51.31$\pm$22.75 & 341.24$\pm$2.40 & 5005.81$\pm$0.47 & -113.22$\pm$50.90 & 774.14$\pm$6.59 \\
J1242+3317 & 41.65 & 5006.86$\pm$0.04 & 1.02$\pm$2.40 & 307.40$\pm$1.20 & 5003.30$\pm$0.10 & -213.29$\pm$6.59 & 888.36$\pm$2.99 \\
J1246+0222 & 41.12 & 5006.03$\pm$0.07 & -48.68$\pm$4.19 & 310.22$\pm$3.59 & 5002.24$\pm$0.49 & -227.11$\pm$31.73 & 738.89$\pm$13.77 \\
J1311+0648 & 41.49 & 5005.42$\pm$0.07 & -85.20$\pm$4.19 & 221.38$\pm$4.19 & 5003.40$\pm$0.21 & -121.05$\pm$14.37 & 671.20$\pm$10.78 \\
J1315+4325 & 40.65 & 5005.79$\pm$0.82 & -63.05$\pm$49.10 & 509.04$\pm$59.28 & 4995.20$\pm$3.12 & -633.88$\pm$234.12 & 619.03$\pm$144.90 \\
J1320+2108 & 40.95 & 5006.66$\pm$0.24 & -10.96$\pm$14.37 & 335.60$\pm$31.14 & 5002.50$\pm$1.16 & -249.42$\pm$82.03 & 934.89$\pm$24.55 \\
J1322+0809 & 40.84 & 5006.10$\pm$0.15 & -44.49$\pm$8.98 & 228.44$\pm$9.58 & 5002.08$\pm$0.50 & -240.60$\pm$37.12 & 619.03$\pm$16.77 \\
J1331+0131 & 41.24 & 5006.92$\pm$0.06 & 4.61$\pm$3.59 & 509.04$\pm$1.80 & 4994.90$\pm$0.53 & -719.78$\pm$34.13 & 337.01$\pm$19.16 \\
J1332+3127 & 41.31 & 5007.55$\pm$0.18 & 42.33$\pm$10.78 & 332.78$\pm$13.77 & 4999.67$\pm$0.68 & -471.48$\pm$50.30 & 905.28$\pm$17.36 \\
J1337+2423 & 42.24 & 5005.74$\pm$0.25 & -66.04$\pm$14.97 & 352.52$\pm$16.17 & 4995.76$\pm$1.29 & -597.81$\pm$90.41 & 2240.64$\pm$25.75 \\
J1342+0505 & 42.17 & 4991.12$\pm$0.21 & -941.44$\pm$12.57 & 652.87$\pm$14.37 & 4990.02$\pm$0.30 & -66.05$\pm$28.74 & 2322.42$\pm$19.16 \\
J1342+4642 & 41.13 & 5005.79$\pm$0.14 & -63.05$\pm$8.38 & 249.59$\pm$9.58 & 5001.91$\pm$0.93 & -232.37$\pm$61.67 & 644.41$\pm$27.54 \\
J1355+5612 & 42.22 & 5007.64$\pm$0.04 & 47.72$\pm$2.40 & 407.52$\pm$1.20 & 5003.83$\pm$0.10 & -228.43$\pm$7.19 & 1013.86$\pm$3.59 \\
J1358+2511 & 41.48 & 5007.41$\pm$0.07 & 33.95$\pm$4.19 & 279.20$\pm$3.59 & 5002.30$\pm$0.14 & -306.09$\pm$10.78 & 884.13$\pm$4.19 \\
J1402+1720 & 40.96 & 5006.00$\pm$0.05 & -50.48$\pm$2.99 & 200.23$\pm$2.40 & 5003.49$\pm$0.20 & -150.30$\pm$13.17 & 844.65$\pm$10.78 \\
J1402+2159 & 41.49 & 5006.18$\pm$0.05 & -39.70$\pm$2.99 & 291.89$\pm$1.80 & 5003.45$\pm$0.13 & -163.17$\pm$8.98 & 762.86$\pm$5.39 \\
J1406+2223 & 41.50 & 5004.95$\pm$0.24 & -113.35$\pm$14.37 & 480.84$\pm$29.94 & 4996.79$\pm$0.52 & -488.83$\pm$43.71 & 822.08$\pm$20.36 \\
J1408+2409 & 41.76 & 5002.06$\pm$0.11 & -286.39$\pm$6.59 & 658.51$\pm$5.99 & 4990.45$\pm$0.60 & -695.89$\pm$41.31 & 1448.17$\pm$19.16 \\
J1439+3923 & 41.38 & 5006.23$\pm$0.07 & -36.70$\pm$4.19 & 352.52$\pm$3.59 & 4997.06$\pm$0.27 & -549.34$\pm$18.56 & 1035.01$\pm$7.19 \\
J1440+6156 & 42.40 & 5006.83$\pm$0.07 & -0.78$\pm$4.19 & 369.44$\pm$4.79 & 4995.38$\pm$0.26 & -685.51$\pm$17.96 & 1081.54$\pm$7.78 \\
J1441+1604 & 41.59 & 5007.63$\pm$0.05 & 47.12$\pm$2.99 & 298.94$\pm$2.40 & 5005.45$\pm$0.12 & -130.65$\pm$7.78 & 772.73$\pm$4.79 \\
J1442+2623 & 41.18 & 5004.69$\pm$0.33 & -128.91$\pm$19.76 & 332.78$\pm$24.55 & 4999.82$\pm$0.98 & -291.43$\pm$76.64 & 864.39$\pm$28.74 \\
J1444+1536 & 41.23 & 5007.79$\pm$0.08 & 56.70$\pm$4.79 & 300.35$\pm$4.19 & 5005.54$\pm$0.14 & -134.24$\pm$11.38 & 611.98$\pm$5.39 \\
J1448+3559 & 41.62 & 5006.80$\pm$0.09 & -2.57$\pm$5.39 & 324.32$\pm$4.19 & 5000.94$\pm$0.28 & -350.60$\pm$19.76 & 731.84$\pm$8.38 \\
J1451+2709 & 41.83 & 5007.01$\pm$0.04 & 10.00$\pm$2.40 & 162.16$\pm$0.60 & 5003.58$\pm$0.10 & -205.65$\pm$6.59 & 894.00$\pm$3.59 \\
J1536+5433* & 41.06 & 5007.23$\pm$0.75 & 23.17$\pm$44.91 & 455.46$\pm$16.17 & $-$ & $-$ & $-$ \\
J1537+4942 & 41.67 & 5006.48$\pm$0.08 & -21.74$\pm$4.79 & 153.70$\pm$6.59 & 5000.77$\pm$0.37 & -341.82$\pm$25.15 & 837.60$\pm$16.17 \\
J1555+1911 & 40.82 & 5005.91$\pm$0.43 & -55.86$\pm$25.75 & 349.70$\pm$4.79 & $-$ & $-$ & $-$ \\
J1559+3501 & 40.58 & 5005.48$\pm$0.07 & -81.61$\pm$4.19 & 293.30$\pm$3.59 & 4999.65$\pm$0.38 & -349.15$\pm$25.15 & 806.57$\pm$14.97 \\
J1605+3239 & 40.89 & 5007.17$\pm$0.55 & 19.58$\pm$32.93 & 211.51$\pm$68.86 & 5004.47$\pm$1.31 & -161.39$\pm$109.57 & 651.46$\pm$59.28 \\
J1627+4736 & 42.31 & 5007.18$\pm$0.07 & 20.18$\pm$4.19 & 401.88$\pm$2.40 & 5001.68$\pm$0.19 & -329.20$\pm$13.77 & 812.21$\pm$4.79 \\
J2140+0025 & 41.19 & 5007.25$\pm$0.15 & 24.37$\pm$8.98 & 184.72$\pm$11.98 & 5004.55$\pm$0.28 & -162.15$\pm$23.95 & 638.77$\pm$16.17 \\
J2219+1207 & 40.89 & 5007.24$\pm$0.08 & 23.77$\pm$4.79 & 225.61$\pm$3.59 & 5001.95$\pm$0.84 & -316.71$\pm$53.29 & 547.12$\pm$24.55 \\
J2254+0046 & 41.03 & 5007.81$\pm$0.25 & 57.90$\pm$14.97 & 370.85$\pm$10.18 & 5000.63$\pm$0.66 & -429.52$\pm$52.69 & 737.48$\pm$25.75 \\
\hline
\end{tabular}
}
\tablefoot{ Columns: (1) short name of the source; (2) logarithm of the [O III] luminosity (\ergs); (3) wavelength of the [O III] core component (\AA{}); (4) shift of the [O III] core with respect to the rest-frame wavelength (\kms); (5) FWHM of the [O III] core component (\kms); (6) wavelength of the [O III] wing component (\AA{}); and (7) velocity of the [O III] wing component with respect to the core (\kms); FWHM of the wing component (\kms). Sources marked with an asterisk are those where the redshift is calculated with respect to H$\beta$ narrow component. }
\end{table*}
\clearpage
\begin{table*}
\caption{Summary of the [O III] line properties for RLNLS1s. }
\label{tab:summary4}
\centering
\scalebox{0.9}{
\begin{tabular}{l c l l l l l l}
\hline\hline
Short Name & $\log{L}_{[O III]}$ & $\lambda_c$ & v$_c$ & FWHM$_c$ & $\lambda_w$ & v$_w$ & FWHM$_w$ \\
\hline\hline
J0138+1321 & 42.02 & 5007.26$\pm$0.10 & 24.97$\pm$5.99 & 407.52$\pm$5.39 & 5003.49$\pm$0.47 & -226.10$\pm$32.33 & 943.35$\pm$14.97 \\
J0146$-$0040 & 41.16 & 5006.67$\pm$0.05 & -10.36$\pm$2.99 & 143.83$\pm$4.19 & 5005.78$\pm$0.22 & -53.34$\pm$14.37 & 575.32$\pm$19.16 \\
J0251$-$0702 & 42.53 & 5005.64$\pm$0.07 & -72.03$\pm$4.19 & 404.70$\pm$5.39 & 5002.05$\pm$0.15 & -214.87$\pm$11.38 & 1116.79$\pm$5.39 \\
J0324+3410* & 40.95 & 5006.71$\pm$1.32 & -7.96$\pm$79.04 & 445.59$\pm$116.76 & 5004.64$\pm$8.16 & -123.73$\pm$565.83 & 1201.40$\pm$341.90 \\
J0706+3901 & 41.33 & 5003.61$\pm$0.07 & -193.58$\pm$4.19 & 736.07$\pm$2.99 & $-$ & $-$ & $-$ \\
J0713+3820 & 42.39 & 5011.86$\pm$0.14 & 300.40$\pm$8.38 & 504.81$\pm$12.57 & 4999.46$\pm$0.58 & -741.81$\pm$41.31 & 1494.70$\pm$17.96 \\
J0804+3853 & 42.21 & 5006.67$\pm$0.04 & -10.36$\pm$2.40 & 169.21$\pm$1.20 & 5001.29$\pm$0.20 & -322.21$\pm$13.17 & 857.34$\pm$7.19 \\
J0806+7248 & 41.81 & 5001.90$\pm$0.20 & -295.97$\pm$11.98 & 375.08$\pm$8.98 & 4994.79$\pm$1.49 & -426.03$\pm$99.40 & 733.25$\pm$61.67 \\
J0814+5609* & 41.73 & 4997.51$\pm$0.51 & -558.83$\pm$30.54 & 740.30$\pm$14.97 & $-$ & $-$ & $-$ \\
J0849+5108 & 41.74 & 5011.26$\pm$0.18 & 264.47$\pm$10.78 & 231.26$\pm$8.98 & 5002.93$\pm$1.09 & -498.77$\pm$74.25 & 703.64$\pm$50.30 \\
J0850+4626 & 42.32 & 5008.52$\pm$0.16 & 100.41$\pm$9.58 & 248.18$\pm$14.97 & 5005.76$\pm$0.52 & -164.70$\pm$38.92 & 1558.15$\pm$28.74 \\
J0902+0443 & 42.20 & 5007.75$\pm$0.27 & 54.31$\pm$16.17 & 286.25$\pm$26.35 & 4999.03$\pm$0.66 & -521.95$\pm$54.49 & 1231.01$\pm$20.36 \\
J0937+3615 & 41.76 & 5006.98$\pm$0.14 & 8.20$\pm$8.38 & 404.70$\pm$11.98 & 5002.80$\pm$0.69 & -249.94$\pm$48.50 & 999.76$\pm$25.15 \\
J0948+0022 & 41.27 & 4993.97$\pm$1.05 & -770.79$\pm$62.87 & 1438.29$\pm$55.69 & $-$ & $-$ & $-$ \\
J0952$-$0136 & 41.67 & 5005.96$\pm$0.08 & -52.87$\pm$4.79 & 558.40$\pm$4.19 & 4994.51$\pm$0.52 & -685.82$\pm$33.53 & 1583.53$\pm$13.77 \\
J0953+2836 & 42.07 & 5006.25$\pm$0.23 & -35.51$\pm$13.77 & 344.06$\pm$16.17 & 4995.81$\pm$1.38 & -624.77$\pm$94.60 & 592.24$\pm$56.88 \\
J1031+4234 & 41.83 & 5009.07$\pm$0.30 & 133.35$\pm$17.96 & 231.26$\pm$10.78 & 5005.45$\pm$0.92 & -216.94$\pm$65.86 & 744.53$\pm$10.18 \\
J1034+3938 & 41.03 & 5006.76$\pm$0.04 & -4.97$\pm$2.40 & 228.44$\pm$1.80 & 5002.54$\pm$0.17 & -253.01$\pm$10.78 & 901.05$\pm$4.79 \\
J1037+0036* & 41.60 & 5008.56$\pm$0.42 & 102.81$\pm$25.15 & 280.61$\pm$7.78 & 4999.72$\pm$0.58 & -529.11$\pm$58.08 & 356.75$\pm$22.15 \\
J1038+4227 & 41.62 & 5005.11$\pm$0.12 & -103.77$\pm$7.19 & 314.45$\pm$7.78 & 4999.87$\pm$0.52 & -313.96$\pm$36.52 & 1146.41$\pm$28.14 \\
J1047+4725 & 42.89 & 5006.66$\pm$0.07 & -10.96$\pm$4.19 & 280.61$\pm$4.79 & 5004.82$\pm$0.39 & -110.43$\pm$25.75 & 844.65$\pm$24.55 \\
J1048+2222 & 41.79 & 5003.35$\pm$0.31 & -209.15$\pm$18.56 & 389.19$\pm$19.76 & 4994.49$\pm$0.75 & -530.88$\pm$61.07 & 1238.06$\pm$16.77 \\
J1102+2239 & 42.38 & 4997.40$\pm$0.36 & -565.41$\pm$21.56 & 879.90$\pm$11.38 & 4984.90$\pm$2.41 & -749.79$\pm$164.06 & 1319.85$\pm$83.23 \\
J1110+3653 & 41.73 & 5007.53$\pm$0.22 & 41.14$\pm$13.17 & 400.47$\pm$16.77 & $-$ & $-$ & $-$ \\
J1133+0432 & 41.42 & 5008.83$\pm$0.04 & 118.97$\pm$2.40 & 235.49$\pm$8.38 & $-$ & $-$ & $-$ \\
J1138+3653 & 41.68 & 5005.39$\pm$0.17 & -87.00$\pm$10.18 & 234.08$\pm$16.17 & 5004.59$\pm$0.43 & -48.40$\pm$34.73 & 923.61$\pm$28.74 \\
J1146+3236 & 42.00 & 5007.10$\pm$0.10 & 15.39$\pm$5.99 & 307.40$\pm$4.19 & 5001.47$\pm$0.40 & -336.86$\pm$28.74 & 620.44$\pm$7.78 \\
J1200$-$0046 & 41.87 & 5007.01$\pm$0.10 & 10.00$\pm$5.99 & 289.07$\pm$17.36 & 5008.26$\pm$0.22 & 74.89$\pm$17.36 & 757.22$\pm$26.35 \\
J1227+3214 & 42.11 & 5007.84$\pm$0.05 & 59.70$\pm$2.99 & 242.54$\pm$1.80 & 5004.35$\pm$0.12 & -209.05$\pm$8.38 & 592.24$\pm$3.59 \\
J1238+3942 & 41.98 & 5005.75$\pm$0.15 & -65.45$\pm$8.98 & 380.73$\pm$15.57 & $-$ & $-$ & $-$ \\
J1246+0238* & 41.64 & 5007.51$\pm$0.35 & 39.94$\pm$20.96 & 382.14$\pm$20.36 & 4999.10$\pm$1.96 & -503.49$\pm$136.52 & 1245.11$\pm$70.06 \\
J1302+1624 & 42.31 & 5003.65$\pm$0.05 & -191.19$\pm$2.99 & 408.93$\pm$1.20 & $-$ & $-$ & $-$ \\
J1305+5116 & 43.14 & 5003.62$\pm$1.36 & -192.98$\pm$81.43 & 1012.45$\pm$77.84 & 4993.35$\pm$2.45 & -615.00$\pm$226.33 & 1394.58$\pm$43.11 \\
J1333+4141 & 42.34 & 5005.35$\pm$0.08 & -89.40$\pm$4.79 & 521.73$\pm$3.59 & 5001.35$\pm$0.11 & -239.37$\pm$10.18 & 1745.69$\pm$5.39 \\
J1337+6005 & 41.75 & 5005.22$\pm$0.11 & -97.18$\pm$6.59 & 355.34$\pm$6.59 & 4997.93$\pm$0.42 & -436.74$\pm$29.94 & 1128.07$\pm$14.97 \\
J1346+3121 & 41.39 & 5005.55$\pm$0.51 & -77.42$\pm$30.54 & 248.18$\pm$55.09 & 5001.72$\pm$2.60 & -229.05$\pm$184.42 & 920.79$\pm$102.39 \\
J1358+2658 & 42.36 & 5007.86$\pm$0.05 & 60.89$\pm$2.99 & 448.41$\pm$3.59 & 5003.64$\pm$0.24 & -252.57$\pm$16.17 & 1366.38$\pm$5.39 \\
J1409+5656 & 41.26 & 5005.27$\pm$0.35 & -94.19$\pm$20.96 & 569.68$\pm$21.56 & $-$ & $-$ & $-$ \\
J1432+3014 & 42.60 & 5007.06$\pm$0.04 & 12.99$\pm$2.40 & 346.88$\pm$1.80 & 5002.79$\pm$0.19 & -255.70$\pm$11.98 & 1145.00$\pm$7.19 \\
J1435+3131 & 42.16 & 5005.74$\pm$0.17 & -66.04$\pm$10.18 & 294.71$\pm$15.57 & 5003.26$\pm$1.01 & -148.30$\pm$69.46 & 1049.11$\pm$97.60 \\
J1443+4725 & 42.32 & 4999.40$\pm$0.54 & -445.66$\pm$32.33 & 1199.99$\pm$33.53 & $-$ & $-$ & $-$ \\
J1450+5919 & 41.55 & 5006.30$\pm$0.06 & -32.51$\pm$3.59 & 265.10$\pm$3.59 & 5002.13$\pm$0.27 & -249.65$\pm$18.56 & 971.55$\pm$9.58 \\
J1505+0326 & 41.77 & 5000.96$\pm$0.99 & -352.25$\pm$59.28 & 597.88$\pm$60.48 & 4995.52$\pm$4.34 & -326.51$\pm$317.35 & 1180.25$\pm$131.13 \\
J1507+4453 & 42.50 & 5007.29$\pm$0.04 & 26.76$\pm$2.40 & 298.94$\pm$1.80 & 5003.18$\pm$0.10 & -245.96$\pm$6.59 & 1187.30$\pm$4.79 \\
J1548+3511 & 42.40 & 5006.63$\pm$0.07 & -12.75$\pm$4.19 & 265.10$\pm$2.40 & 4999.95$\pm$0.59 & -400.03$\pm$37.72 & 971.55$\pm$27.54 \\
J1608+0708 & 41.17 & 5006.31$\pm$0.14 & -31.91$\pm$8.38 & 279.20$\pm$9.58 & 5000.14$\pm$0.60 & -369.45$\pm$42.51 & 1016.68$\pm$21.56 \\
J1612+4219 & 42.31 & 5004.23$\pm$0.14 & -156.46$\pm$8.38 & 741.71$\pm$12.57 & 5005.96$\pm$0.36 & 103.66$\pm$28.14 & 1913.50$\pm$20.96 \\
J1629+4007 & 42.06 & 5009.08$\pm$0.06 & 133.94$\pm$3.59 & 211.51$\pm$2.40 & 5004.52$\pm$0.34 & -272.60$\pm$22.15 & 830.54$\pm$19.16 \\
J1633+4718 & 41.65 & 5006.57$\pm$0.05 & -16.35$\pm$2.99 & 255.23$\pm$1.80 & 5001.91$\pm$0.23 & -278.77$\pm$14.97 & 1185.89$\pm$13.17 \\
J1634+4809 & 41.96 & 5002.47$\pm$0.44 & -261.84$\pm$26.35 & 389.19$\pm$24.55 & 4993.44$\pm$1.69 & -541.39$\pm$125.14 & 1109.74$\pm$56.28 \\
J1644+2619* & 40.91 & 5006.68$\pm$0.44 & -9.76$\pm$26.35 & 145.24$\pm$9.58 & $-$ & $-$ & $-$ \\
J1703+4540 & 41.55 & 5010.34$\pm$1.23 & 209.39$\pm$73.65 & 335.60$\pm$49.10 & 5004.73$\pm$3.96 & -335.68$\pm$309.56 & 645.82$\pm$110.77 \\
J1709+2348 & 42.14 & 5008.43$\pm$0.09 & 95.02$\pm$5.39 & 401.88$\pm$3.59 & 5003.06$\pm$0.39 & -321.53$\pm$26.94 & 786.83$\pm$9.58 \\
J1713+3523 & 40.98 & 4995.59$\pm$0.88 & -673.79$\pm$52.69 & 1641.35$\pm$47.90 & $-$ & $-$ & $-$ \\
J1722+5654 & 42.19 & 5006.64$\pm$0.12 & -12.15$\pm$7.19 & 321.50$\pm$5.99 & 4999.30$\pm$0.45 & -439.04$\pm$32.33 & 716.33$\pm$21.56 \\
J2314+2243* & 42.31 & 5006.86$\pm$0.81 & 1.02$\pm$48.50 & 602.11$\pm$25.75 & 4989.64$\pm$2.20 & -1031.42$\pm$178.43 & 1579.30$\pm$55.69 \\
\hline
\end{tabular}
}
\tablefoot{Columns as in Table~\ref{tab:summary3}.}
\end{table*}

\end{document}